\documentclass[10pt]{article}

\usepackage[T1]{fontenc}
\usepackage{lmodern}
\usepackage{microtype}

\usepackage[letterpaper,margin=0.8in]{geometry} 

\usepackage{amsmath,amssymb}
\usepackage{mathrsfs} 

\usepackage{graphicx}
\usepackage[table]{xcolor}
\usepackage{longtable}
\usepackage{multirow}
\usepackage[export]{adjustbox}

\usepackage{caption}
\usepackage{subcaption}

\usepackage{authblk}
\usepackage{etoolbox}     
\usepackage{orcidlink}

\usepackage{textcomp}

\usepackage[numbers,sort&compress]{natbib} 

\usepackage{hyperref}
\hypersetup{colorlinks=true, linkcolor=blue, citecolor=blue, urlcolor=blue}

\begin{document}

\begin{center}
\LARGE{Perturbations and Greybody Factors of AdS Black Holes with a Cloud of Strings Surrounded by Quintessence-like Field in NLED Scenario}
\end{center}

\vspace{0.1cm}

\begin{center}

{\bf Faizuddin Ahmed}\orcidlink{0000-0003-2196-9622}\\
Department of Physics, The Assam Royal Global University, Guwahati, 781035, Assam, India\\
e-mail: faizuddinahmed15@gmail.com\\

\vspace{0.15cm}

{\bf Ahmad Al-Badawi}\orcidlink{0000-0002-3127-3453}\\
Department of Physics, Al-Hussein Bin Talal University, 71111,
Ma'an, Jordan. \\
e-mail: ahmadbadawi@ahu.edu.jo\\

\vspace{0.15cm}

{\bf \.{I}zzet Sakall{\i}}\orcidlink{0000-0001-7827-9476}\\
Physics Department, Eastern Mediterranean University, Famagusta 99628, North Cyprus via Mersin 10, Turkey\\
e-mail: izzet.sakalli@emu.edu.tr

\author{\textbf{Sara Kanzi}\orcidlink{0000-0001-7181-2813}\footnote{\textbf{Email:} sara.kanzi@final.edu.tr}\\}
\textit{Faculty of Engineering, Final International University, North Cyprus via Mersin 10, Kyrenia 99320, Turkey}

\vspace{0.15cm}

{\bf Sara Kanzi}\orcidlink{0000-0001-7181-2813}\\
Faculty of Engineering, Final International University, North Cyprus via Mersin 10, Kyrenia 99320, Turkey.\\
e-mail: sara.kanzi@final.edu.tr

\end{center}

\vspace{0.2cm}

\begin{abstract}
The discovery of gravitational waves and advances in black hole imaging have opened unprecedented opportunities to probe exotic physics in the strong-field regime. Building upon a recent black hole solution in Einstein gravity coupled with nonlinear electrodynamics and exotic matter sources—specifically a cloud of strings and quintessence field—we investigate the perturbative dynamics, thermodynamic properties, and transmission characteristics of quantum fields in anti-de Sitter spacetime. The black hole is characterized by mass, nonlinear magnetic charge, string cloud parameter, and quintessence field parameters, which collectively modify the spacetime geometry, horizon structure, Hawking temperature, and effective potentials governing field propagation. We derive Schr\"{o}dinger-like wave equations for massless scalar, electromagnetic, and fermionic perturbations, analyzing how these multiple matter sources simultaneously influence the potential barrier structures. Our Hawking temperature analysis reveals striking dependencies on both the event horizon radius and nonlinear charge parameter, displaying behavior fundamentally distinct from asymptotically flat black holes due to the cosmological constant. We calculate greybody factors—quantifying Hawking radiation transmission probabilities—for all field spins using turning point analysis. A striking discovery emerges in the fermionic sector: positive and negative helicity modes exhibit optimal transmission at different quintessence normalization values, demonstrating explicit helicity-dependent coupling absent in bosonic channels. This asymmetry provides a unique signature of spin-exotic matter interactions with potential implications for detecting quintessence through precision measurements of black hole radiation spectra. Our results extend previous relevant perturbation studies by incorporating coupled effects of nonlinear electrodynamics, cosmic strings, and quintessence, offering new outcomes relevant to gravitational wave astronomy and early universe cosmology.
\end{abstract}


{\color{black}
\section{Introduction} \label{isec1}

Black holes (BHs) represent among the most fascinating predictions of Einstein's general relativity, serving as natural laboratories for exploring the interplay between quantum mechanics, thermodynamics, and gravitational physics \cite{ref1,ref2}. The discovery of gravitational waves by the LIGO-Virgo-KAGRA collaboration has ushered in a new era of observational astronomy, enabling direct detection of BH mergers and providing unprecedented opportunities to test fundamental physics in the strong-field regime \cite{ref3,ref4,ref5}. A particularly important aspect of BH physics concerns the response of these objects to external perturbations, which manifests through the dynamics of field propagation in curved spacetime backgrounds \cite{ref6,ref38}. The study of such perturbations reveals fundamental information about the BH's intrinsic properties—mass, charge, angular momentum, and potentially parameters associated with exotic matter fields or modified gravity theories \cite{ref9,ref10}.

Anti-de Sitter (AdS) spacetimes, characterized by a negative cosmological constant $\Lambda < 0$, have attracted considerable attention due to their relevance in the AdS/CFT correspondence, which establishes a profound duality between gravitational theories in AdS space and conformal field theories on the boundary \cite{ref11,ref12,ref13}. This holographic principle has profound implications for quantum gravity, black hole information paradox, and strongly coupled gauge theories \cite{ref14,ref15}. Beyond their theoretical significance, AdS BHs exhibit rich phenomenology, including stable photon spheres, modified thermodynamic properties, and distinct perturbative behavior compared to their asymptotically flat counterparts \cite{ref16,ref17,ref18}.

The incorporation of exotic matter sources into BH solutions provides valuable insights into scenarios relevant to early universe cosmology, dark energy models, and modified gravity theories. Two prominent examples are cloud of strings (CS) and quintessence fields (QF). String clouds, first introduced by Letelier \cite{ref19}, arise naturally in cosmological models involving cosmic strings and topological defects formed during phase transitions in the early universe \cite{ref20,ref21}. The energy-momentum tensor of a CS exhibits a unique structure with radial pressure balancing energy density, leading to significant modifications of spacetime geometry and BH horizon structures \cite{ref22,ref23}. Quintessence fields, on the other hand, represent a class of dynamical dark energy models characterized by equation-of-state parameters $-1 < w < -1/3$, which have been extensively studied in cosmology to explain the observed accelerated expansion of the universe \cite{ref24,ref25,ref26}. When incorporated into BH solutions via the Kiselev ansatz \cite{ref24}, quintessence fields introduce power-law modifications to the metric functions, altering thermodynamic properties, geodesic structures, and stability characteristics \cite{ref27,ref28,ref29}.

Nonlinear electrodynamics (NLED) provides an alternative framework for describing electromagnetic phenomena in strong-field regimes where quantum corrections and nonlinear interactions become significant \cite{ref30,ref31,ref32}. Unlike Maxwell's linear electrodynamics, NLED theories predict modified dispersion relations, light-cone structures, and charge-dependent metric functions that can regularize classical singularities or introduce novel causal structures \cite{ref33,ref34}. The exponential NLED model, characterized by the charge coupling $e^{-k/r}$ in the metric function (where $k = Q^2/(2M)$), has been particularly successful in constructing regular or weakly singular BH solutions while preserving asymptotic flatness or AdS behavior at large distances \cite{ref35,ref36,ref37}.

The study of BH perturbations encompasses interconnected physical phenomena that together provide a comprehensive framework for understanding BH stability, thermodynamic properties, and observational signatures. Perturbation dynamics are governed by Schrödinger-like wave equations with effective potentials encoding spacetime geometry and field spin \cite{ref6,ref38}. Greybody factors (GBFs) quantify the transmission probability of Hawking radiation through effective potential barriers and measure deviations from ideal blackbody spectra \cite{ref41,ref42,ref43,ref44}. Recent investigations have explored perturbations in various modified BH scenarios, including Gauss-Bonnet gravity \cite{ref45}, Horndeski theories \cite{ref46}, Einstein-Maxwell-dilaton systems \cite{ref47}, and BHs surrounded by exotic matter fields \cite{ref27,ref48,ref49}. However, the combined effects of NLED, CS, and QF on perturbative dynamics in AdS backgrounds remain relatively unexplored, particularly concerning spin-dependent transmission characteristics and helicity asymmetries in fermionic channels.

Driven by the phenomenology resulting from the interaction of NLED, exotic matter sources, and negative cosmological constant, the present work undertakes a comprehensive investigation of field perturbations and GBFs in the AdS-NLED-CS-QF BH solution recently derived by Nascimento {\it et al.} \cite{ref35}. Our first objective is to derive and analyze the effective potentials governing massless scalar (spin-0), electromagnetic (spin-1), and fermionic (spin-1/2) field perturbations in the AdS-NLED-CS-QF background. By reducing the field equations to Schrödinger-like forms, we systematically investigate how the parameters $\{\alpha, k, N, w, \Lambda\}$ modify the potential barrier structures, examining their dependence on angular momentum quantum numbers $\ell$ and field spin. This analysis extends previous studies \cite{ref51,ref52,ref53,ref54} by incorporating the simultaneous presence of multiple exotic matter sources within the NLED framework. Our second objective is to investigate the transmission coefficients or the so-called greybody factors (GBFs) for scalar, electromagnetic, and fermionic fields using the turning point approximation. A particularly novel aspect of our study is the identification of helicity-dependent transmission characteristics in the fermionic sector, where positive and negative helicity modes exhibit optimal transmission at different quintessence normalization values ($N = 0.5$ versus $N = 0.1$). This helicity asymmetry, absent in bosonic channels, represents a unique signature of spin-exotic matter coupling that could have implications for fermionic Hawking radiation spectra.

Our findings contribute to the broader understanding of how exotic matter fields and NLED modify BH observables, with potential relevance to constraining dark energy models through BH perturbative signatures \cite{ref25,ref26}, testing NLED theories in astrophysical contexts \cite{ref32,ref36}, exploring quantum field dynamics in curved spacetimes with multiple matter sources \cite{ref48,ref49}, and developing observational strategies for detecting exotic matter effects via gravitational wave observations and precision BH measurements. The results presented here may also inform theoretical investigations of early universe phase transitions involving cosmic strings and quintessence-dominated epochs, as well as provide testable predictions for future high-precision gravitational wave detectors and very-long-baseline interferometry observations of supermassive BH shadows.

The remainder of this paper is organized as follows. Section~\ref{isec2} reviews the AdS-NLED-CS-QF BH solution, presenting the metric function, energy-momentum tensors for CS and QF, and analyzing the horizon structure across parameter space. Section~\ref{isec3} derives the perturbation equations for spin-0 scalar, spin-1 electromagnetic, and spin-1/2 fermionic fields, reducing them to Schrödinger-like wave equations and examining the effective potentials in detail. Section~\ref{isec5} investigates the GBFs for all field spins, analyzing transmission dynamics and highlighting the remarkable helicity-dependent behavior in the fermionic sector. Finally, Sec.~\ref{isec6} summarizes our main conclusions and discusses future research directions.
}

\section{AdS-NLED BH Solution with CS and QF: Geometry and Horizons} \label{isec2}

The incorporation of exotic matter sources into black hole solutions within general relativity has been a subject of sustained theoretical interest, particularly in exploring how such fields modify classical spacetime geometries and thermodynamic properties. Two prominent examples of such exotic sources are CS and QF, both of which play significant roles in cosmological scenarios and modified gravity theories.

The pioneering study of a black hole solution with a CS as the matter source was presented by Letelier \cite{ref19}. In that seminal work, he derived a generalization of the Schwarzschild BH surrounded by a spherically symmetric CS, characterized by the energy-momentum tensor:
\begin{equation}
    T^{t}_{t}=T^{r}_{r}=\rho_c=\frac{\alpha}{r^2},\quad T^{\theta}_{\theta}=T^{\phi}_{\phi}=0,\label{pp1}
\end{equation}
where $\rho_c$ denotes the energy density of the string cloud and $\alpha$ is an integration constant quantifying the string cloud's strength. The string cloud parameter $\alpha$ must satisfy $0 \leq \alpha < 1$ to ensure physically meaningful solutions. The Letelier BH metric is given by \cite{ref19}
\begin{equation}
    ds^2=-\left(1-\alpha-\frac{2\,M}{r}\right)\,dt^2+\left(1-\alpha-\frac{2\,M}{r}\right)^{-1}\,dr^2+r^2\,(d\theta^2+\sin^2 \theta\,d\phi^2).\label{pp2}
\end{equation}
This solution demonstrates that the presence of a string cloud effectively reduces the effective mass as seen at large distances, leading to modified horizon structures and altered causal properties compared to the standard Schwarzschild geometry \cite{ref57}.

Parallel to the CS investigations, the study of quintessence fields as matter content within general relativity was pioneered by Kiselev \cite{ref24}, who obtained a generalization of the Schwarzschild solution describing a BH surrounded by a quintessence field. The QF is characterized by the energy-momentum tensor:
\begin{equation}
    T^{t}_{t}=T^{r}_{r}=\rho_q,\quad T^{\theta}_{\theta}=T^{\phi}_{\phi}=-\frac{1}{2}\,\rho_q\,(3\,w_q+1),\label{pp3}
\end{equation}
where $\rho_q$ denotes the energy density of the quintessence field, and the pressure is related to the density via the equation of state $p_q = w_q \rho_q$. Here, $w_q$ is the quintessence state parameter, which must satisfy $-1 < w_q < -1/3$ to describe accelerated cosmic expansion consistent with dark energy models \cite{ref24,ref15}. The corresponding Kiselev metric is given by:
\begin{equation}
    ds^2=-\left(1-\frac{2\,M}{r}-\frac{q}{r^{3\,w_q+1}}\right)\,dt^2+\left(1-\frac{2\,M}{r}-\frac{q}{r^{3\,w_q+1}}\right)^{-1}\,dr^2+r^2\,(d\theta^2+\sin^2 \theta\,d\phi^2).\label{pp4}
\end{equation}

Building upon these foundational works, we now consider the scenario where both the quintessence field and the CS coexist in the spacetime, assuming they do not interact with each other. This non-interaction hypothesis allows us to treat their combined energy-momentum tensor as a linear superposition of the individual contributions:
\begin{equation}
    T^{t}_{t}=T^{r}_{r}=\rho_q+\frac{\alpha}{r^2},\quad T^{\theta}_{\theta}=T^{\phi}_{\phi}=-\frac{1}{2}\,\rho_q\,(3\,w_q+1).\label{pp5}
\end{equation}
This superposition principle is justified when the coupling between the two exotic matter fields is negligible, a reasonable approximation in many cosmological contexts \cite{ref59}.

The static and spherically symmetric AdS black hole solution incorporating both quintessence and a CS within the framework of NLED was recently derived by do Nascimento {\it et al.} \cite{ref35}. The line element is given by:
\begin{eqnarray}
ds^2=-f(r)\,dt^2+\frac{dr^2}{f(r)}+r^2\,\left(d\theta^2+\sin ^2 \theta \,d \phi^2\right),\label{aa1}
\end{eqnarray}
where the metric function $f(r)$ encodes the gravitational effects of the BH mass, NLED charge, CS, QF, and the negative cosmological constant:
\begin{eqnarray}
f(r)=1-\alpha-\frac{2\,M}{r}\,e^{-k/r}-\frac{\mathrm{N}}{r^{3\,w+1}}-\frac{\Lambda}{3}\,r^2.\label{aa2}
\end{eqnarray}

To maintain consistency with the notation used in Ref.~\cite{ref35} and throughout the remainder of this manuscript, we adopt the following conventions for the quintessence field parameters. The quintessence equation-of-state parameter is denoted by $w$ (without subscript) rather than $w_q$, where $-1 < w < -1/3$ characterizes the quintessence matter. Similarly, the quintessence normalization constant is represented by $\mathrm{N}$ instead of the alternative notation $q$ or $c$ found in some literature. These notational choices, while equivalent to those in Eqs.~(\ref{pp3})--(\ref{pp5}), streamline the presentation and avoid potential confusion with the magnetic charge $Q$ and other parameters. Explicitly, the correspondence is: $w \equiv w_q$ and $\mathrm{N} \equiv q \equiv c$.

In Eq.~(\ref{aa2}), $0 \leq \alpha <1$ represents the cosmic string parameter, $M$ denotes the BH mass, and $k=\frac{Q^2}{2\,M}$ encodes the nonlinear self-gravitating magnetic charge parameter, where $Q$ is the total magnetic charge of the BH. Finally, $\Lambda<0$ is the cosmological constant, which is negative for anti-de Sitter (AdS) spacetime.

The NLED contribution arises from considering a spherically symmetric spacetime sourced by a magnetic field. For such configurations, the only non-vanishing component of the electromagnetic field tensor $F_{\mu\nu}$ is $F_{23}=F_{\theta\phi}=Q\,\sin \theta$, and the electromagnetic invariant is $F=\frac{1}{4}\,F^{\mu\nu}\,F_{\mu\nu} = \frac{Q^2}{2\,r^4}$. The exponential factor $e^{-k/r}$ in Eq.~(\ref{aa2}) captures the nonlinear electromagnetic effects, distinguishing this solution from the standard Reissner-Nordström (RN) case \cite{ref35,ref60}.

An important limiting behavior emerges at large radial distances. For $r \to \infty$, we can expand the exponential term:
\begin{equation}
    e^{-k/r} \approx 1-\frac{k}{r}+\mathcal{O}(1/r^2).\label{aa3}
\end{equation}
Substituting this approximation into Eq.~(\ref{aa2}) and using $k=\frac{Q^2}{2\,M}$, the metric function reduces to:
\begin{equation}
    f(r)=1-\alpha-\frac{2\,M}{r}+\frac{Q^2}{r^2}-\frac{\mathrm{N}}{r^{3\,w+1}}-\frac{\Lambda}{3}\,r^2.\label{aa4}
\end{equation}
This asymptotic form demonstrates that, at large distances, our solution recovers the structure of a spherically symmetric RN-AdS BH metric modified by the presence of a CS and quintessence. The $Q^2/r^2$ term represents the standard Coulomb-like electromagnetic contribution, while the string cloud effectively shifts the mass term by the constant $\alpha$, and the quintessence introduces a power-law decay term $\mathrm{N}/r^{3w+1}$ that dominates at different radial scales depending on the value of $w$.

The event horizons of the BH are determined by solving the equation $f(r_h) = 0$, which generally yields multiple roots depending on the parameter values $\{M, \alpha, k, \mathrm{N}, w, \Lambda\}$. Table~\ref{tab:horizons_M_alpha_k} provides a comprehensive numerical analysis of the horizon structure for various combinations of the mass $M$, cosmic string parameter $\alpha$, and nonlinear charge parameter $k$, while keeping $\Lambda = -0.001$, $\mathrm{N} = 0.01$, and $w = -2/3$ fixed.

\setlength{\tabcolsep}{10pt}
\renewcommand{\arraystretch}{1.3}
\begin{longtable}{|c|c|c|c|c|}
\hline
\rowcolor{brown!50}
\textbf{$M$} & \textbf{$\alpha$} & \textbf{$k$} & \textbf{Horizon(s)} & \textbf{Configuration} \\
\hline
\endfirsthead


1.0 & 0.0 & 0.0 & $[2.0387400]$ & Extremal or Single Root BH \\
\hline
1.0 & 0.0 & 0.1 & $[0.022222021,\ 1.9342272]$ & Non-extremal BH \\
\hline
1.0 & 0.0 & 0.5 & $[0.23174037,\ 1.4289552]$ & Non-extremal BH \\
\hline
1.0 & 0.1 & 0.0 & $[2.2753855]$ & Extremal or Single Root BH \\
\hline
1.0 & 0.1 & 0.1 & $[0.021575160,\ 2.1707413]$ & Non-extremal BH \\
\hline
1.0 & 0.1 & 0.5 & $[0.21301490,\ 1.6796895]$ & Non-extremal BH \\
\hline
1.0 & 0.2 & 0.0 & $[2.5758144]$ & Extremal or Single Root BH \\
\hline
1.0 & 0.2 & 0.1 & $[0.020900724,\ 2.4708707]$ & Non-extremal BH \\
\hline
1.0 & 0.2 & 0.5 & $[0.19633537,\ 1.9910921]$ & Non-extremal BH \\
\hline

\caption{\footnotesize Horizons $r_h$ for various values of the mass $M$, cosmic string parameter $\alpha$, and nonlinear charge parameter $k$. 
The constants are fixed as $\Lambda = -0.001$, $\mathrm{N} = 0.01$, and $w = -2/3$. 
The mass is set to $M = 1.0$, $\alpha \in \{0.0,\,0.1,\,0.2\}$, and $k \in \{0.0,\,0.1,\,0.5\}$, 
where $k = Q^{2}/(2M)$ corresponds to the nonlinear charge parameter (e.g., for $M=1.0$ and $Q=0.5$, one obtains $k=0.125$). 
Each configuration is categorized as extremal (single horizon) or non-extremal (two distinct horizons, corresponding to inner Cauchy and outer event horizons).}
\label{tab:horizons_M_alpha_k}
\end{longtable}

\begin{figure}[ht!]
    \centering
    \includegraphics[width=0.75\linewidth]{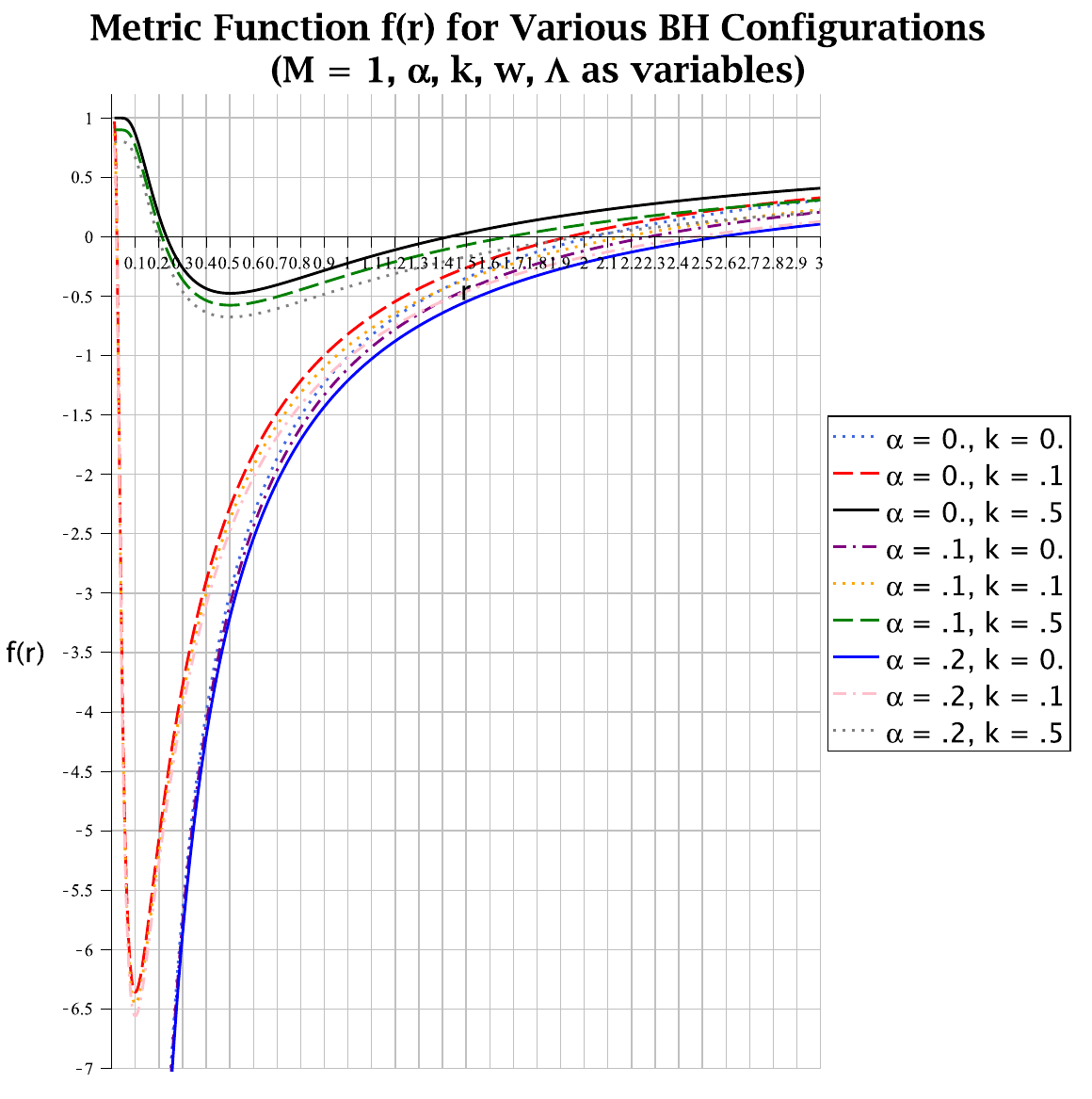}
    \caption{\footnotesize Metric function $f(r)$ as a function of radial coordinate $r$ for various BH configurations characterized by different CS parameter $\alpha$ and NLED charge parameter $k$ values. Parameters: $M = 1$, $w = -2/3$, $\Lambda = -0.001$, $\mathrm{N} = 0.01$. The zeros of $f(r)$ correspond to horizon locations listed in Table~\ref{tab:horizons_M_alpha_k}. Configurations with $k = 0$ exhibit single horizons (extremal), while $k > 0$ produces two distinct horizons (inner Cauchy and outer event horizons). Increasing $\alpha$ shifts the metric function downward, reflecting weakened effective gravity due to the CS. All curves asymptotically approach $-\infty$ as $r \to \infty$ due to the negative cosmological constant characteristic of AdS spacetime. The legend indicates the specific $(\alpha, k)$ values for each curve.}
    \label{fig:metric_function}
\end{figure}

Several key observations emerge from Table~\ref{tab:horizons_M_alpha_k}. First, when the NLED parameter $k=0$ (corresponding to zero electromagnetic charge), the BH possesses only a single horizon, indicating an extremal or near-extremal configuration. The introduction of nonzero charge ($k > 0$) generically leads to the formation of two distinct horizons: an inner Cauchy horizon and an outer event horizon, characteristic of charged BH solutions. Second, increasing the cosmic string parameter $\alpha$ systematically enlarges both horizons, shifting them to larger radial coordinates. This effect can be understood as a consequence of the string cloud reducing the effective gravitational attraction, thereby expanding the trapped region. Third, for fixed $\alpha$, increasing the NLED charge parameter $k$ brings the inner and outer horizons closer together, eventually merging at a critical value corresponding to an extremal BH configuration. Beyond this critical charge, no horizons exist, and the spacetime contains a naked singularity, violating cosmic censorship \cite{ref61}. These horizon structure results are crucial for understanding the thermodynamic stability, Hawking radiation properties, and perturbative dynamics of the AdS-NLED-CS-QF BH system, which we investigate in the subsequent sections of this work. To visualize the behavior of the metric function across different parameter regimes, Fig.~\ref{fig:metric_function} displays $f(r)$ as a function of radial coordinate $r$ for the nine parameter combinations presented in Table~\ref{tab:horizons_M_alpha_k}. The plot provides valuable insights into how the CS parameter $\alpha$ and NLED charge parameter $k$ collectively influence the spacetime geometry. Several key features are evident from this analysis. First, the zeros of the metric function $f(r) = 0$ correspond precisely to the horizon locations tabulated in Table~\ref{tab:horizons_M_alpha_k}, confirming the numerical accuracy of our horizon calculations. For configurations with $k = 0$ (corresponding to uncharged BHs), represented by the blue, purple, and blue curves for $\alpha = 0.0, 0.1, 0.2$ respectively, only a single horizon exists, manifesting as a single zero crossing in the positive $r$ region. These extremal or near-extremal configurations exhibit monotonic behavior of $f(r)$ with no secondary horizon structure. Second, introducing nonzero NLED charge ($k > 0$) fundamentally alters the metric topology by creating two distinct horizons: an inner Cauchy horizon at small $r$ and an outer event horizon at larger $r$. This bifurcation is clearly visible in the red, black, orange, green, pink, and gray curves, where $f(r)$ crosses zero twice. The separation between these horizons decreases systematically as $k$ increases from $0.1$ to $0.5$ for fixed $\alpha$, indicating that stronger NLED effects bring the inner and outer horizons closer together. Third, increasing the CS parameter $\alpha$ shifts the entire metric function downward (toward more negative values at small $r$), reflecting the reduction in effective gravitational attraction induced by the string cloud. This downward shift is most pronounced in the asymptotic regime $r \to 0$, where $f(r)$ plunges to increasingly negative values for larger $\alpha$. The physical interpretation is that the CS effectively weakens the gravitational field, requiring larger radial coordinates to achieve the same metric function value compared to configurations with smaller $\alpha$. Fourth, in the large-$r$ regime, all curves eventually become negative due to the dominant $-\Lambda r^2/3$ term arising from the negative cosmological constant characterizing AdS spacetime. This asymptotic behavior is universal across all parameter combinations and represents the characteristic "AdS throat" structure where the metric function diverges to $-\infty$ as $r \to \infty$. The competition between the repulsive cosmological constant term and the attractive gravitational and NLED terms determines the location of the outer event horizon. Fifth, the curves exhibit varying degrees of curvature in the intermediate region between horizons, reflecting the interplay among the mass term $2M e^{-k/r}/r$, the NLED exponential coupling, the QF contribution $\mathrm{N}/r^{3w+1}$, and the CS parameter $\alpha$. The smoother curves (e.g., black curve for $\alpha = 0.0, k = 0.5$) correspond to configurations where NLED effects significantly modify the near-horizon geometry, while steeper curves indicate parameter regimes where gravitational attraction dominates. This visualization complements the numerical horizon data in Table~\ref{tab:horizons_M_alpha_k} and provides intuitive understanding of how the various physical parameters sculpt the spacetime geometry of the AdS-NLED-CS-QF BH system.

To further elucidate the thermodynamic properties of the AdS-NLED-CS-QF BH system, we investigate the Hawking temperature as a function of the event horizon radius $r_h$ and the NLED charge parameter $k$. The Hawking temperature of a BH is determined by the surface gravity at the event horizon through the fundamental relation
\begin{equation}
T_H = \frac{\kappa}{2\pi},\label{hawking_temp}
\end{equation}
where $\kappa$ is the surface gravity defined as
\begin{equation}
\kappa = \frac{1}{2}\left.\frac{df(r)}{dr}\right|_{r=r_h} = \frac{1}{2}f'(r_h).\label{surface_gravity}
\end{equation}
For the metric function given in Eq.~(\ref{aa2}), the derivative with respect to the radial coordinate is
\begin{equation}
f'(r) = \frac{2M}{r^2}\,e^{-k/r} + \frac{kM}{r^3}\,e^{-k/r} + \frac{\mathrm{N}(3w+1)}{r^{3w+2}} - \frac{2\Lambda}{3}.\label{metric_derivative}
\end{equation}
Evaluating this at the event horizon $r = r_h$ (where $f(r_h) = 0$) and substituting into Eqs.~(\ref{surface_gravity}) and (\ref{hawking_temp}), we obtain the Hawking temperature
\begin{equation}
T_H(r_h, k) = \frac{1}{4\pi}\left[\frac{2M}{r_h^2}\,e^{-k/r_h} + \frac{kM}{r_h^3}\,e^{-k/r_h} + \frac{\mathrm{N}(3w+1)}{r_h^{3w+2}} - \frac{2\Lambda}{3}\right].\label{hawking_temp_explicit}
\end{equation}
This expression encodes the combined contributions from the BH mass, NLED charge, quintessence field, and the negative cosmological constant to the thermal radiation emitted by the BH.

\begin{figure}[ht!]
    \centering
    \includegraphics[width=0.65\linewidth]{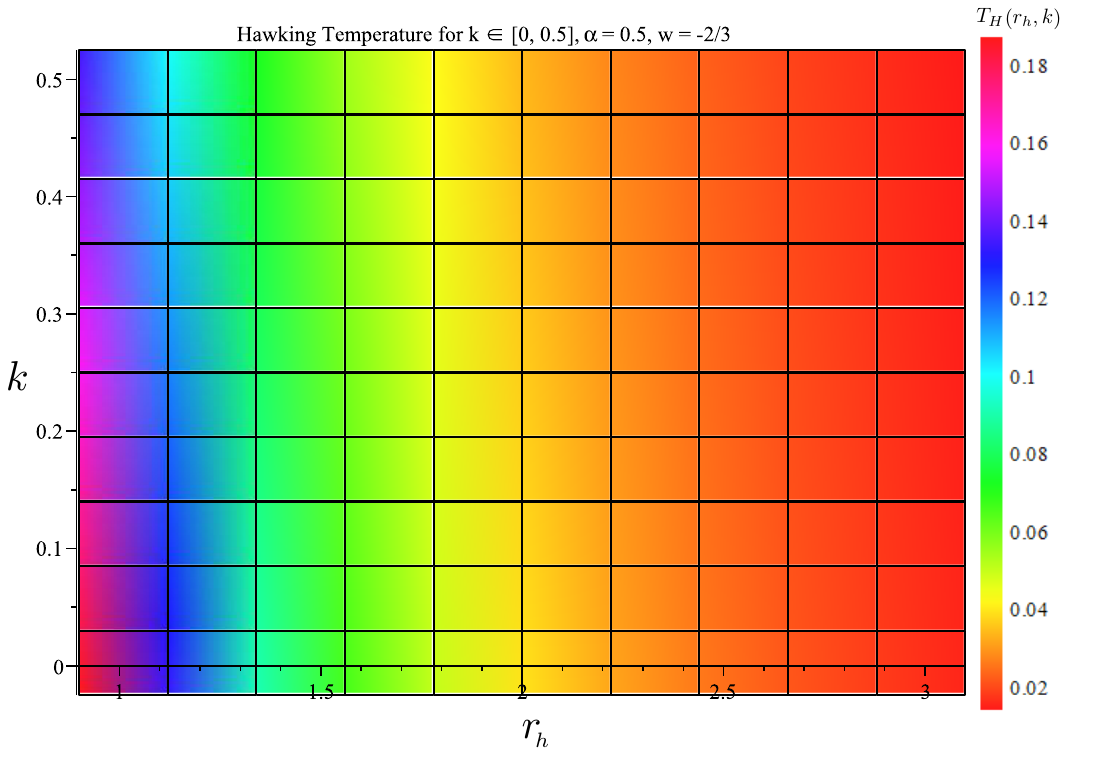}
   \caption{\footnotesize Density plot of Hawking temperature $T_H$ as a function of event horizon radius $r_h$ and NLED charge parameter $k$ for the AdS-NLED-CS-QF BH. Parameters: $M = 1$, $\alpha = 0.2$, $w = -2/3$, $\Lambda = -0.001$, $\mathrm{N} = 0.01$. The color bar indicates temperature values (in units where $G = c = \hbar = k_B = 1$), with blue representing lower temperatures and red representing higher temperatures. The plot demonstrates that Hawking temperature increases with both horizon radius and NLED charge parameter, reflecting the interplay between AdS geometry, NLED, CS, and QF contributions to the surface gravity.}
\label{fig:hawking_temp}
\end{figure}

As an ilustrative example, Fig.~\ref{fig:hawking_temp} presents a density plot of the Hawking temperature in the $(r_h, k)$ parameter space for fixed values $\alpha = 0.2$, $w = -2/3$, $\Lambda = -0.001$, and $\mathrm{N} = 0.01$. The color gradient from blue (low temperature) to red (high temperature) reveals several important features of the thermodynamic landscape. First, for fixed $k$, the Hawking temperature increases monotonically with the event horizon radius $r_h$, indicating that larger BHs in this parameter regime are thermodynamically hotter. This behavior differs markedly from asymptotically flat Schwarzschild BHs, where larger horizons correspond to cooler temperatures ($T_H \propto 1/r_h$), and reflects the dominant influence of the negative cosmological constant in the AdS geometry. The $-2\Lambda r/3$ term in Eq.~(\ref{hawking_temp_explicit}) provides a positive contribution that grows linearly with $r_h$, eventually dominating over the mass term that decays as $1/r_h^2$. Second, for fixed $r_h$, increasing the NLED charge parameter $k$ leads to a systematic enhancement of the Hawking temperature, as evidenced by the rightward shift toward warmer colors (yellow-orange-red) at larger $k$ values. This temperature amplification arises from the exponential NLED coupling $e^{-k/r}$ in the metric function, which modifies the near-horizon geometry and strengthens the surface gravity through the additional term $kM e^{-k/r_h}/r_h^3$ in Eq.~(\ref{hawking_temp_explicit}). Third, the combined effects of CS parameter $\alpha = 0.2$ and quintessence normalization $\mathrm{N} = 0.01$ introduce additional subtle modifications to the temperature profile. While the CS parameter $\alpha$ does not appear explicitly in the temperature formula (as it only shifts the metric function by a constant), it indirectly affects $T_H$ by modifying the location of the event horizon $r_h$ obtained from solving $f(r_h) = 0$. The quintessence contribution, proportional to $\mathrm{N}(3w+1)/r_h^{3w+2}$, provides a positive contribution to the temperature for the chosen value $w = -2/3$ (since $3w+1 = -1$), though it remains subdominant compared to the NLED and cosmological constant terms in this parameter range. These thermodynamic characteristics have important implications for BH evaporation timescales, thermal stability, and the interplay between Hawking radiation and the surrounding exotic matter fields, which we explore further through perturbative analysis and GBF calculations in subsequent sections.

\section{Field Perturbations and Effective Potentials} \label{isec3}

Perturbation theory constitutes a fundamental framework in gravitational physics, playing a central role in stability analysis of compact objects and the investigation of field dynamics in curved spacetimes \cite{ref59,ref63}. When a BH is subjected to external perturbations—whether from infalling matter, gravitational radiation, or quantum field fluctuations—it responds through dynamics that encode information about its intrinsic properties. These perturbations are typically described by second-order differential equations with appropriate boundary conditions (ingoing waves at the event horizon and outgoing waves at spatial infinity) \cite{ref6,ref38}.

The systematic study of integer-spin perturbations in BH spacetimes has demonstrated that the equations of motion for scalar, electromagnetic, and gravitational fields can be elegantly reduced to Schrödinger-like forms with effective potentials determined by spacetime geometry and field spin \cite{ref6,ref66,ref67}. In this section, we focus on deriving and analyzing the perturbation equations for spin-0 scalar fields, spin-1 electromagnetic fields, and spin-1/2 fermionic fields propagating in the AdS-NLED-CS-QF BH background characterized by the metric function in Eq.~(\ref{aa2}). Our analysis reveals how the simultaneous presence of NLED, CS, and QF modifies the effective potential barriers, thereby influencing field propagation, stability properties, and transmission characteristics.

\subsection{Scalar Field Dynamics: Spin-0 Perturbations}

We begin by examining the dynamics of a massless scalar field propagating in the AdS-NLED-CS-QF BH background. Scalar perturbations are particularly valuable for probing BH stability and have been extensively investigated in various BH geometries, providing fundamental insights into the behavior of test fields and the response of spacetime to external disturbances \cite{ref54,ref52,ref51,ref53}. The evolution of a massless scalar field is governed by the Klein-Gordon equation:
\begin{equation}
\frac{1}{\sqrt{-g}}\,\partial_{\mu}\left[\left(\sqrt{-g}\,g^{\mu\nu}\right)\,\partial_{\nu}\Psi\right]=0,\label{ff1}    
\end{equation}
where $\Psi$ denotes the scalar field wave function, $g_{\mu\nu}$ is the covariant metric tensor, $g=\det(g_{\mu\nu})$ is the metric determinant, $g^{\mu\nu}$ is the contravariant metric tensor, and $\partial_{\mu}$ represents the coordinate derivative. To solve this equation in the spherically symmetric BH background, we employ the standard separation of variables ansatz:
\begin{equation}
    \Psi(t, r,\theta, \phi)=\exp(i\,\omega\,t)\,Y^{m}_{\ell} (\theta,\phi)\,\frac{\psi(r)}{r},\label{ff2}
\end{equation}
where $\omega$ is the temporal frequency characterizing the field oscillation, $\psi(r)$ is the radial wave function, and $Y^{m}_{\ell}(\theta,\phi)$ are the scalar spherical harmonics labeled by angular momentum quantum number $\ell$ and azimuthal quantum number $m$.

Substituting the ansatz (\ref{ff2}) into the Klein-Gordon equation (\ref{ff1}) and performing the standard manipulations, we obtain a one-dimensional radial wave equation. To cast this equation into the canonical Schrödinger form, we introduce the tortoise coordinate $r_*$ defined by:
\begin{eqnarray}
    dr_*=\frac{dr}{f(r)},\label{ff4}
\end{eqnarray}
which transforms the radial coordinate such that the event horizon at $f(r_h) = 0$ is mapped to $r_* \to -\infty$. In terms of the tortoise coordinate, the radial equation becomes:
\begin{equation}
    \frac{\partial^2 \psi(r_*)}{\partial r^2_{*}}+\left(\omega^2-\mathcal{V}_\text{scalar}\right)\,\psi(r_*)=0,\label{ff3}
\end{equation}
where the effective potential for scalar perturbations is given by:
\begin{eqnarray} 
\mathcal{V}_\text{scalar}&=&\left(\frac{\ell\,(\ell+1)}{r^2}+\frac{f'(r)}{r}\right)\,f(r)\nonumber\\
&=&\left[\frac{\ell\,(\ell+1)}{r^2}+\frac{2\,M\,(r-k)}{r^4}\,e^{-k/r}+\frac{\mathrm{N}\,(3\,w+1)}{r^{3\,w+3}}-\frac{2\,\Lambda}{3}\right]\,\left(1-\alpha-\frac{2\,M}{r}\,e^{-k/r}-\frac{\mathrm{N}}{r^{3\,w+1}}-\frac{\Lambda}{3}\,r^2\right).\label{ff5}
\end{eqnarray}

The expression (\ref{ff5}) reveals the intricate structure of the scalar perturbative potential, which encodes contributions from multiple physical sources. The first term in the square brackets, $\ell(\ell+1)/r^2$, represents the familiar centrifugal barrier arising from angular momentum conservation. The second term, proportional to $e^{-k/r}$, captures the NLED corrections to the gravitational potential. The third term, involving the quintessence parameter $\mathrm{N}$ and equation-of-state parameter $w$, reflects the influence of the QF. The fourth term, proportional to $\Lambda$, arises from the negative cosmological constant characterizing AdS spacetime. The second factor in Eq.~(\ref{ff5}) is simply the metric function $f(r)$, which modulates the entire potential and ensures proper asymptotic behavior. Together, these contributions demonstrate how the parameters $\{\alpha, k, \mathrm{N}, w, \Lambda, M, \ell\}$ collectively shape the scalar field dynamics in the AdS-NLED-CS-QF background.

\begin{figure}[ht!]
    \includegraphics[width=0.33\linewidth]{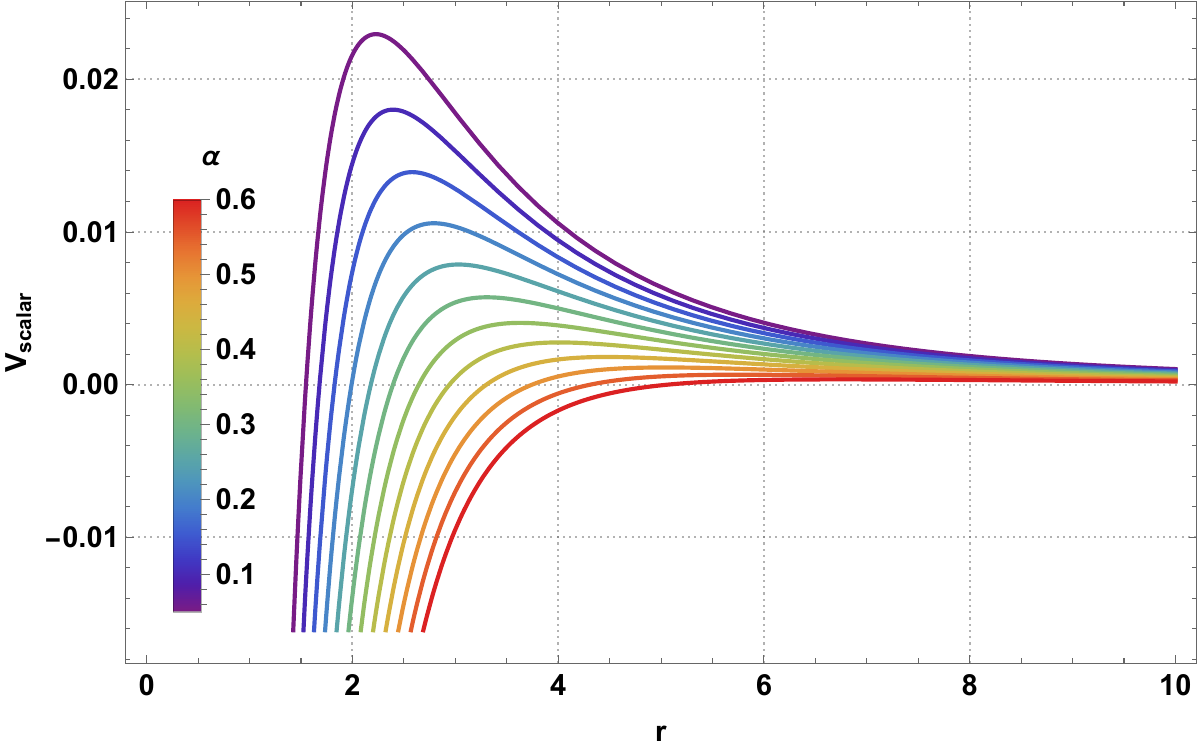}
    \includegraphics[width=0.33\linewidth]{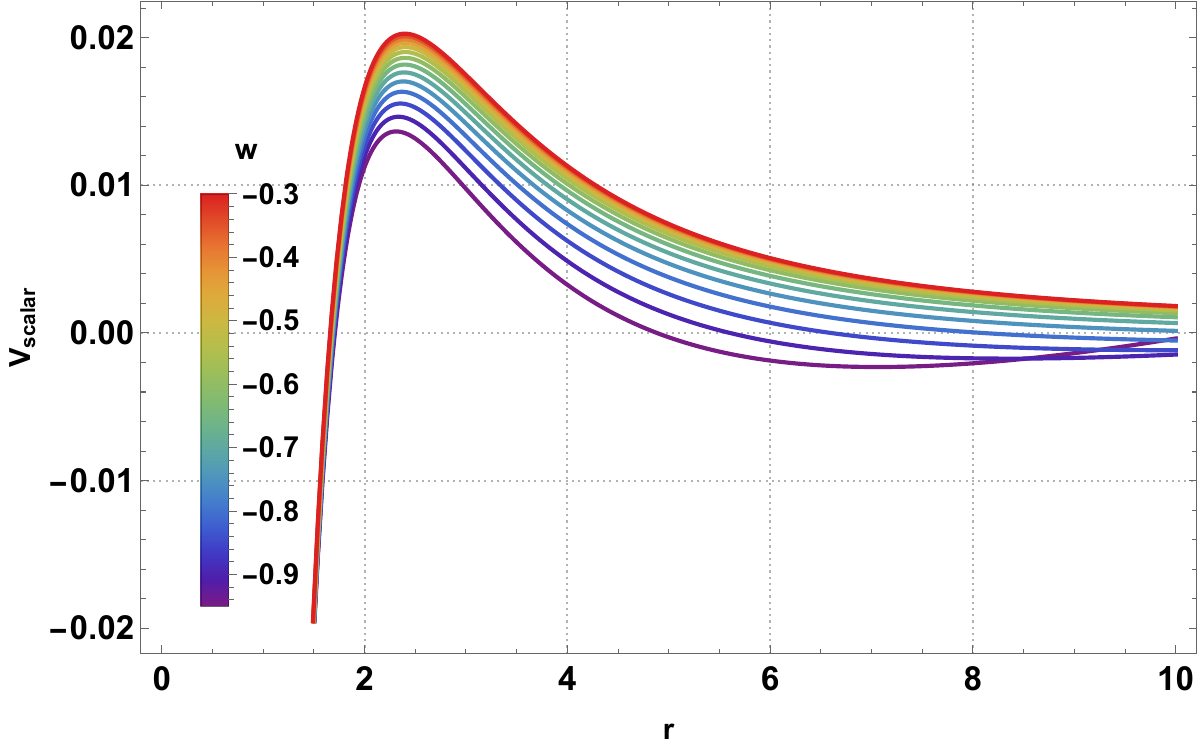}
    \includegraphics[width=0.33\linewidth]{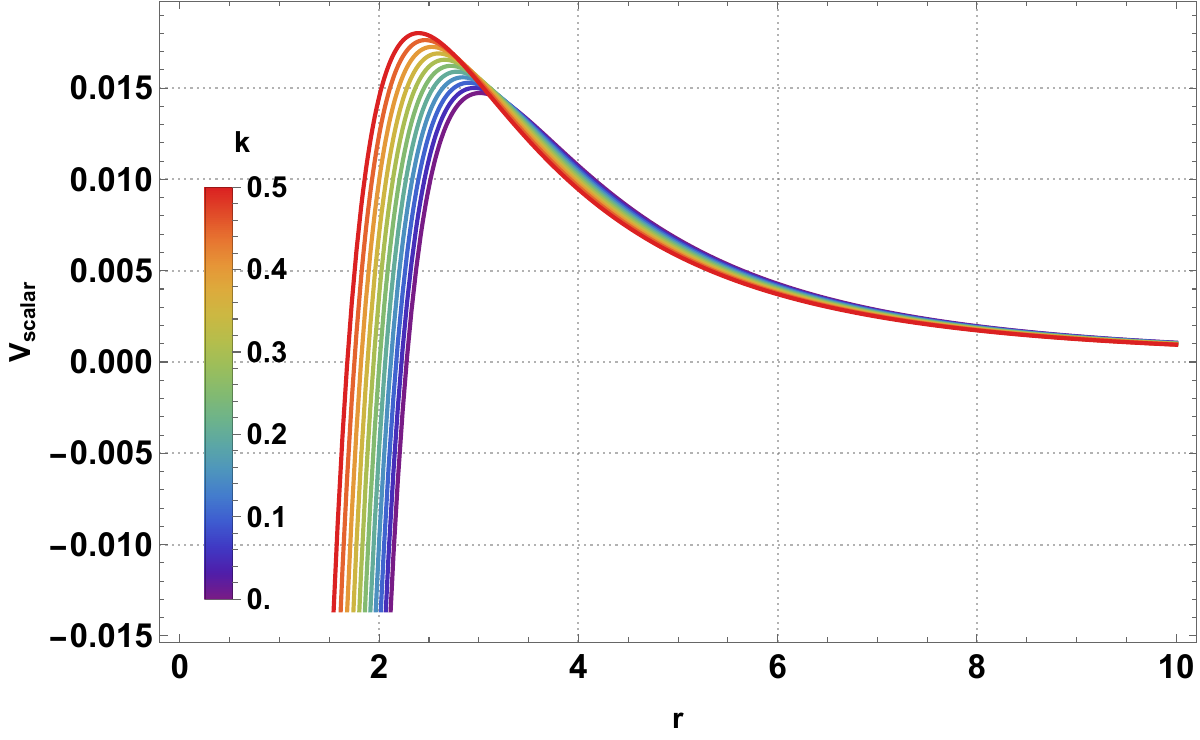}\\
    (i) $Q=1,\,w=-2/3$ \hspace{2cm} (ii) $\alpha=0.1,\,k=0.5$ \hspace{2cm} (iii) $\alpha=0.1,\,w=-2/3$
    \caption{\footnotesize Behavior of the scalar perturbative potential $\mathcal{V}_\text{scalar}$ for $\ell=0$-state by varying values of $\alpha, w$ and $k$. Here $M=1,\,\Lambda=-0.001,\,\mathrm{N}=0.01$. Panel (i) shows the effect of varying CS parameter $\alpha$ while keeping $Q=1$ and $w=-2/3$ fixed. Panel (ii) demonstrates the influence of quintessence parameter $w$ with fixed $\alpha=0.1$ and $k=0.5$. Panel (iii) illustrates the impact of NLED charge parameter $k$ with $\alpha=0.1$ and $w=-2/3$.}
    \label{fig:1}
\end{figure}

\begin{figure}[ht!]
    \centering
    \includegraphics[width=0.325\linewidth]{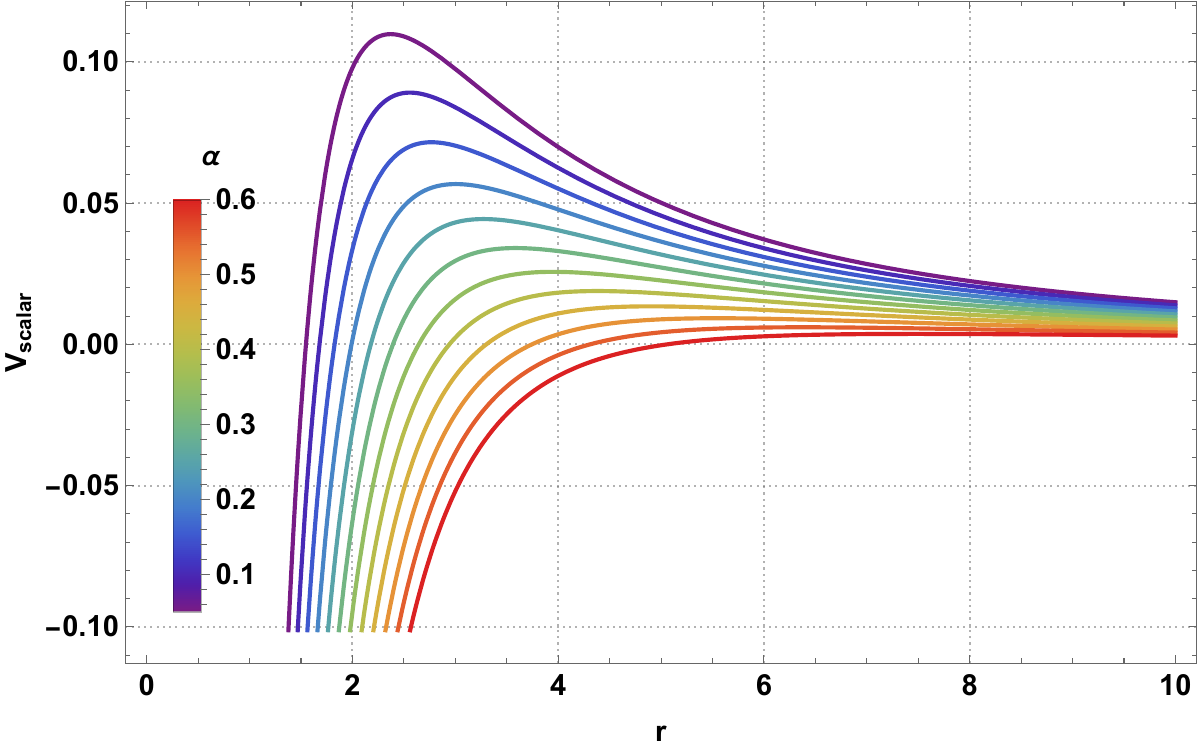}
    \includegraphics[width=0.325\linewidth]{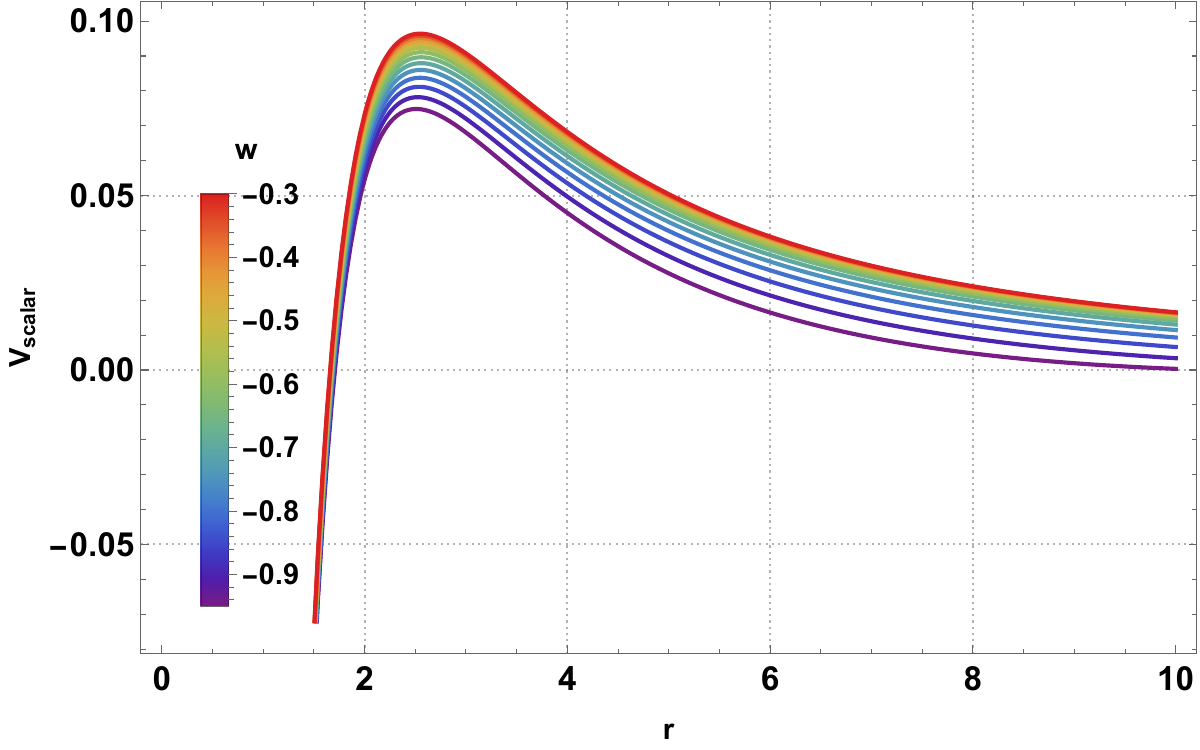}
    \includegraphics[width=0.325\linewidth]{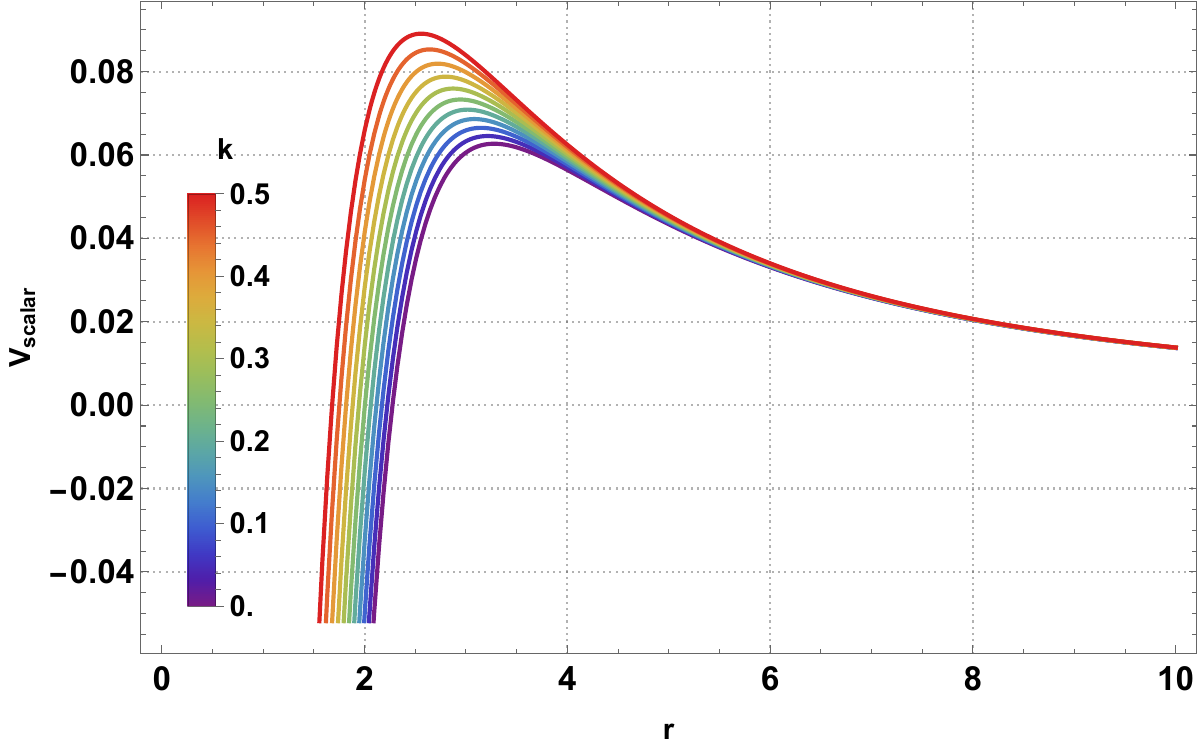}\\
    (i) $Q=1,\,w=-2/3$ \hspace{2cm} (ii) $\alpha=0.1,\,k=0.5$ \hspace{2cm} (iii) $\alpha=0.1,\,w=-2/3$
    \caption{\footnotesize Behavior of the scalar perturbative potential $\mathcal{V}_\text{scalar}$ for $\ell=1$-state by varying values of $\alpha, w$ and $k$. The dipole mode exhibits a pronounced centrifugal barrier compared to the monopole case in Fig.~\ref{fig:1}, with potential peaks occurring at larger radii and achieving higher maximum values.}
    \label{fig:2}
\end{figure}

\begin{figure}[ht!]
    \centering
    \includegraphics[width=0.325\linewidth]{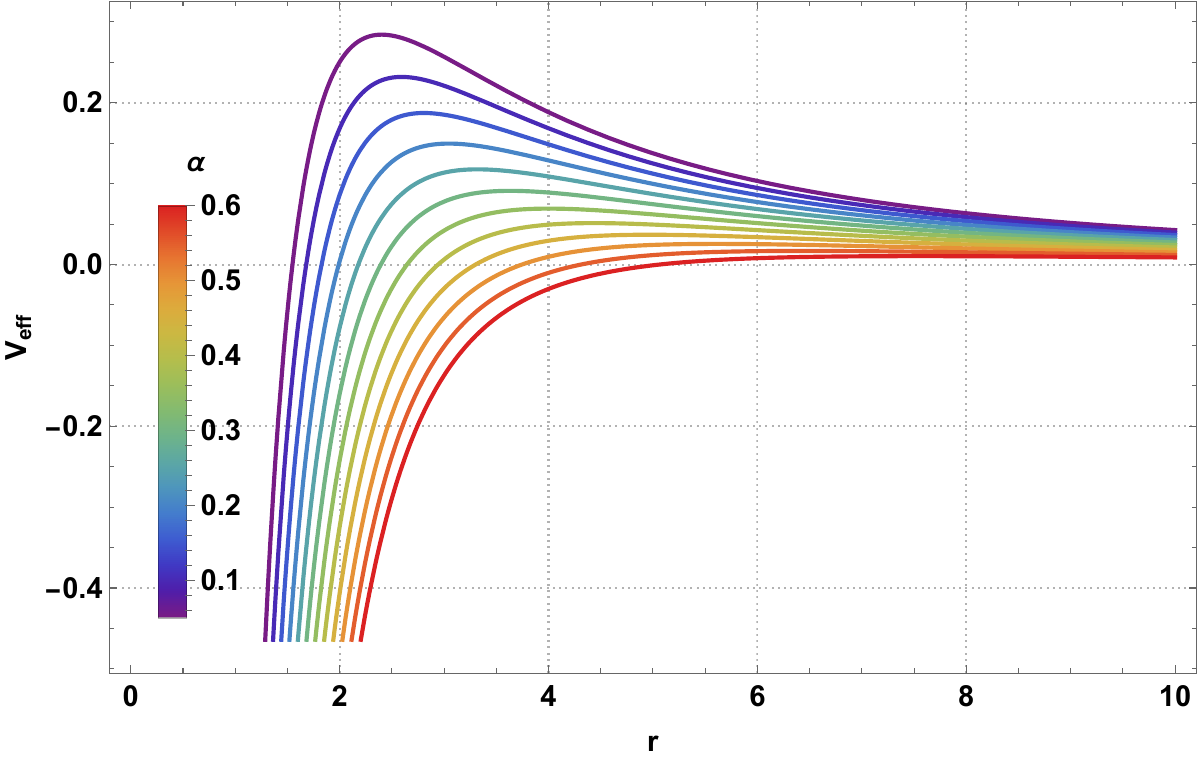}
    \includegraphics[width=0.325\linewidth]{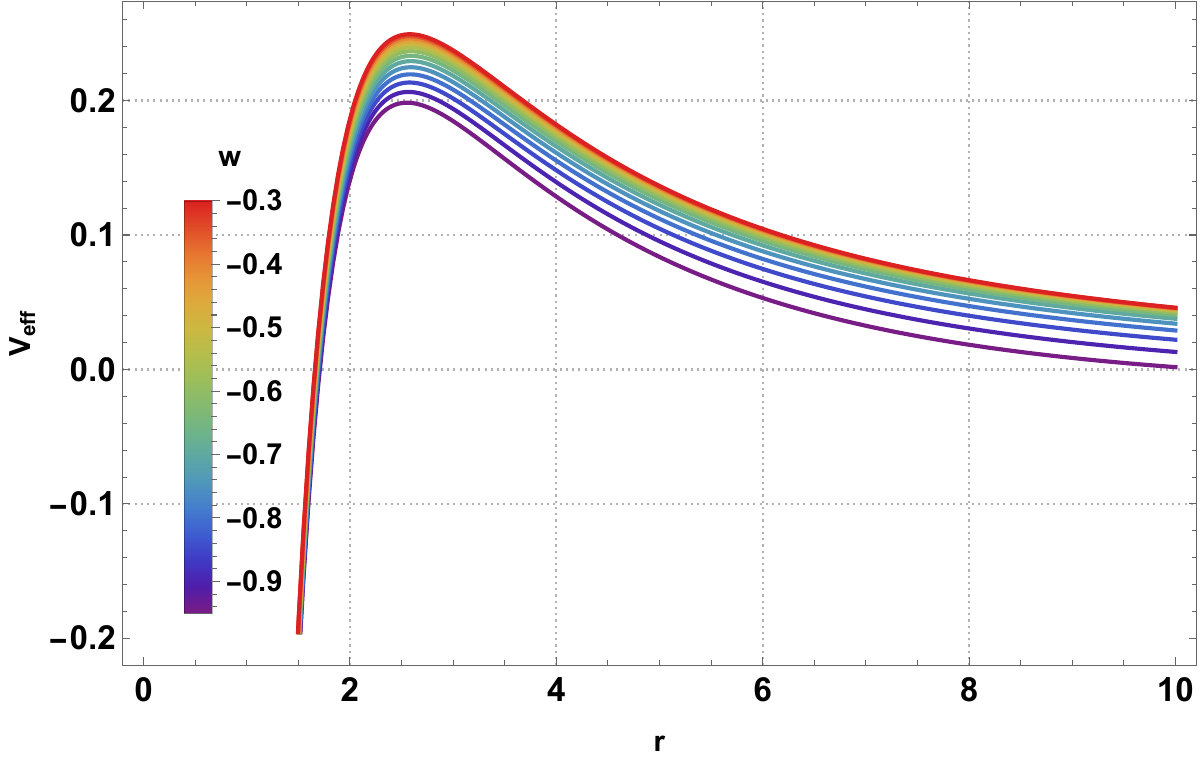}
    \includegraphics[width=0.325\linewidth]{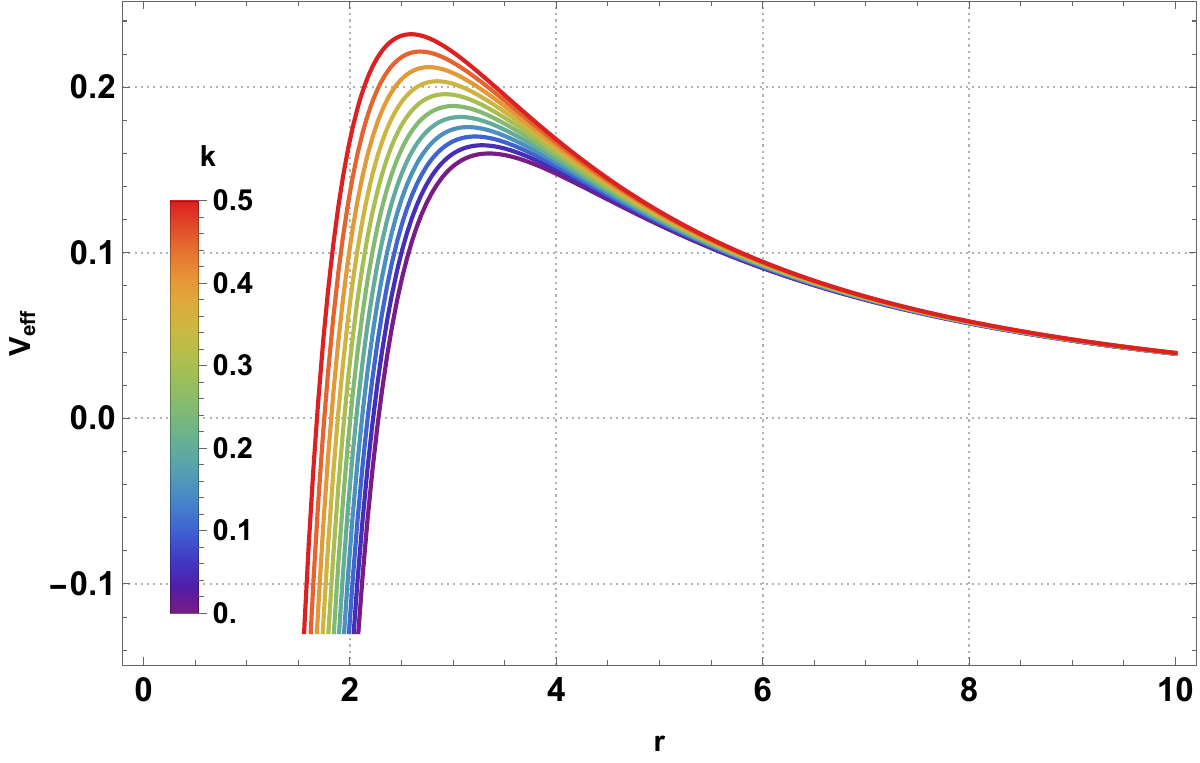}\\
    (i) $Q=1,\,w=-2/3$ \hspace{3cm} (ii) $\alpha=0.1,\,k=0.5$ \hspace{3cm} (iii) $\alpha=0.1,\,w=-2/3$
    \caption{\footnotesize Behavior of the scalar perturbative potential $\mathcal{V}_\text{scalar}$ for $\ell=2$-state by varying values of $\alpha, w$ and $k$.}
    \label{fig:3}
\end{figure}

\begin{figure}[ht!]
    \centering
    \includegraphics[width=0.45\linewidth]{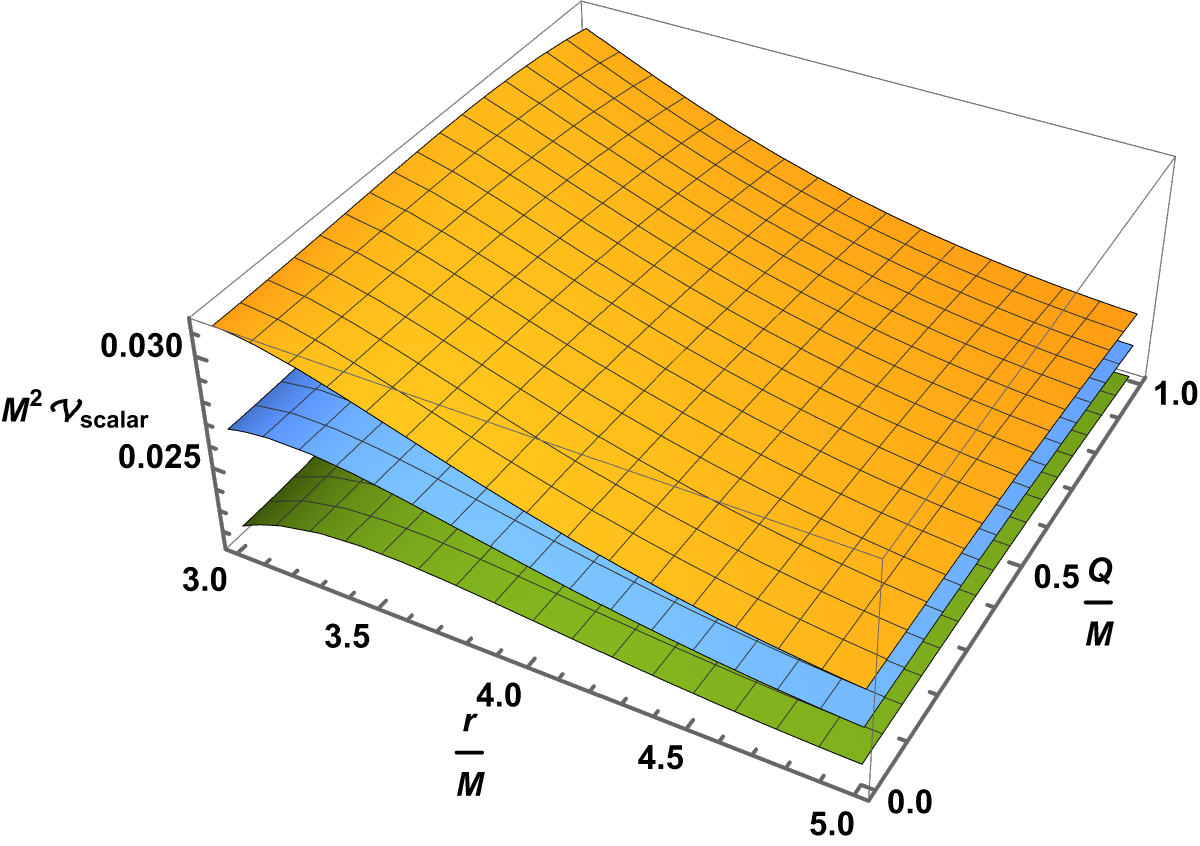}\qquad
    \includegraphics[width=0.45\linewidth]{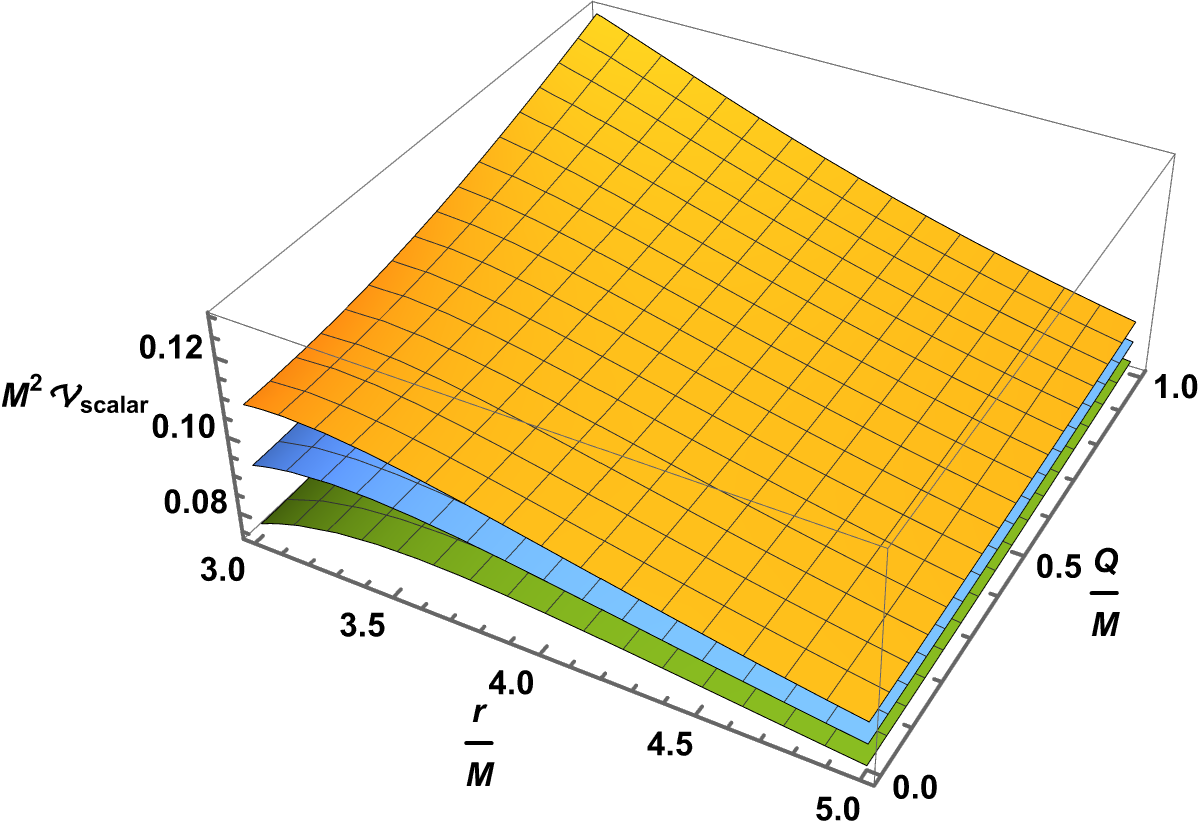}\\
    (i) $\ell=0$ \hspace{8cm} (ii) $\ell=1$
    \caption{\footnotesize Three-dimensional plot of $M^2\,\mathcal{V}_\text{scalar}$ as a function of $(r/M, Q/M)$. Here $\eta=0.1,\,\zeta=0.01$. Yellow $\to \alpha=0.05$; Blue $\to \alpha=0.1$; Green $\to \alpha=0.15$. The layered surfaces illustrate how increasing CS parameter $\alpha$ progressively lowers the effective potential barrier across the entire parameter space.}
    \label{fig:4}
\end{figure}

\begin{figure}[ht!]
    \centering
    \includegraphics[width=0.4\linewidth]{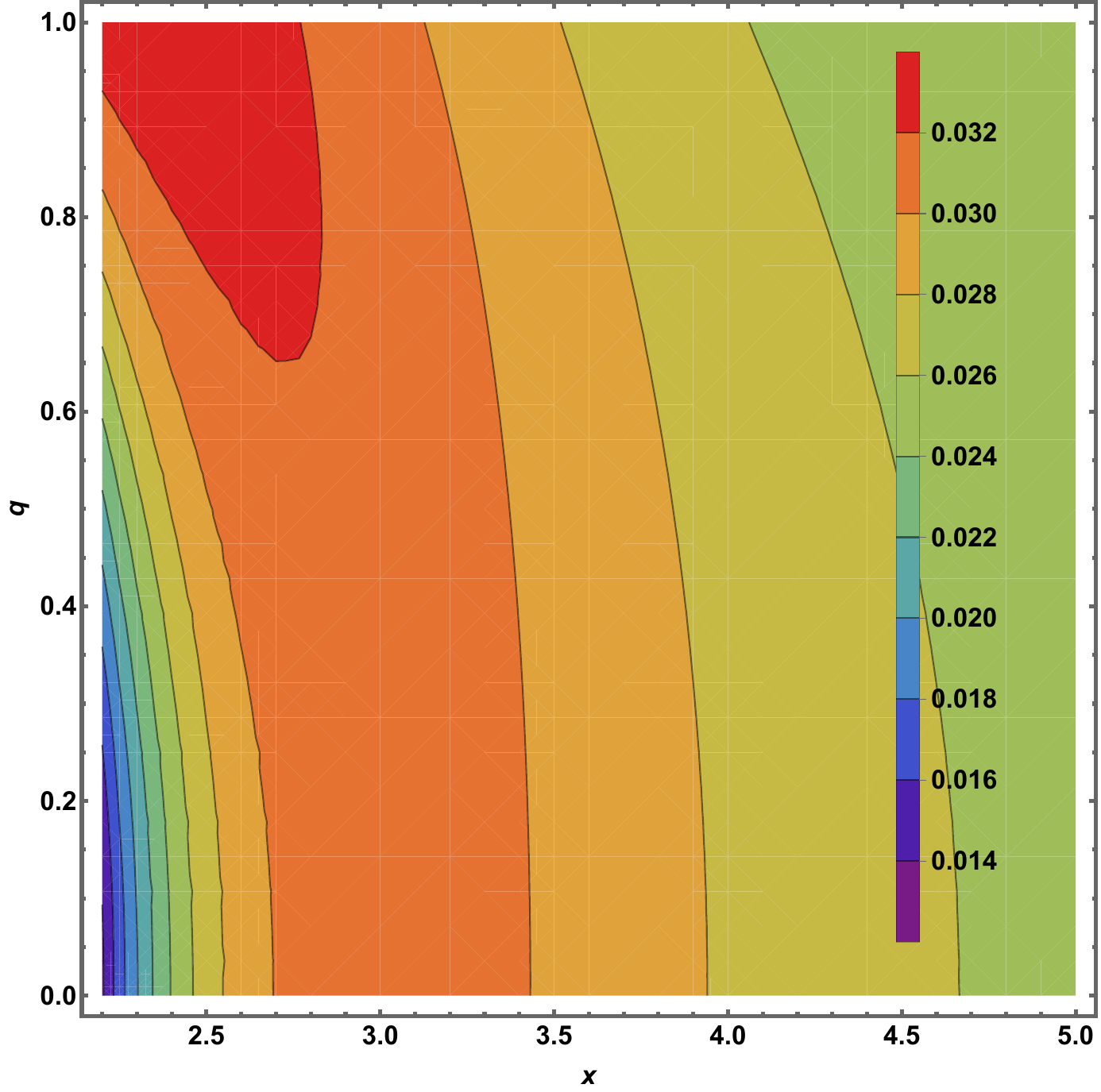}\qquad
    \includegraphics[width=0.4\linewidth]{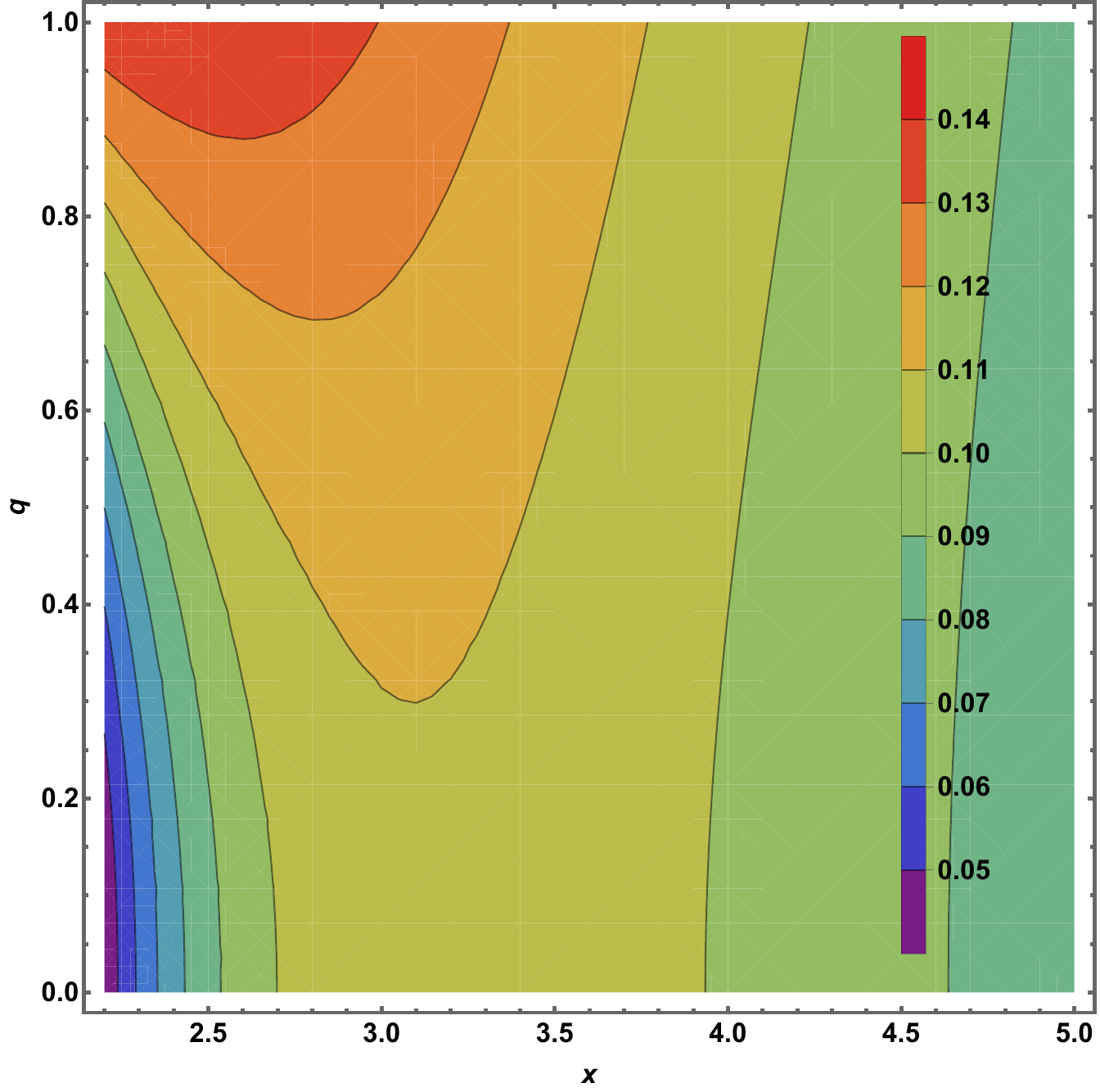}\\
    (i) $\ell=0$ \hspace{8cm} (ii) $\ell=1$
    \caption{Contour plot of $M^2\,\mathcal{V}_\text{scalar}$ for $\alpha=0.05$. Here $\eta=0.1,\,\zeta=0.01$. The contour lines reveal the potential gradient structure, with tighter spacing indicating steeper potential variations in specific regions of the $(x,q)$ parameter space.}
    \label{fig:5}
\end{figure}

\begin{figure}[ht!]
    \centering
    \includegraphics[width=0.4\linewidth]{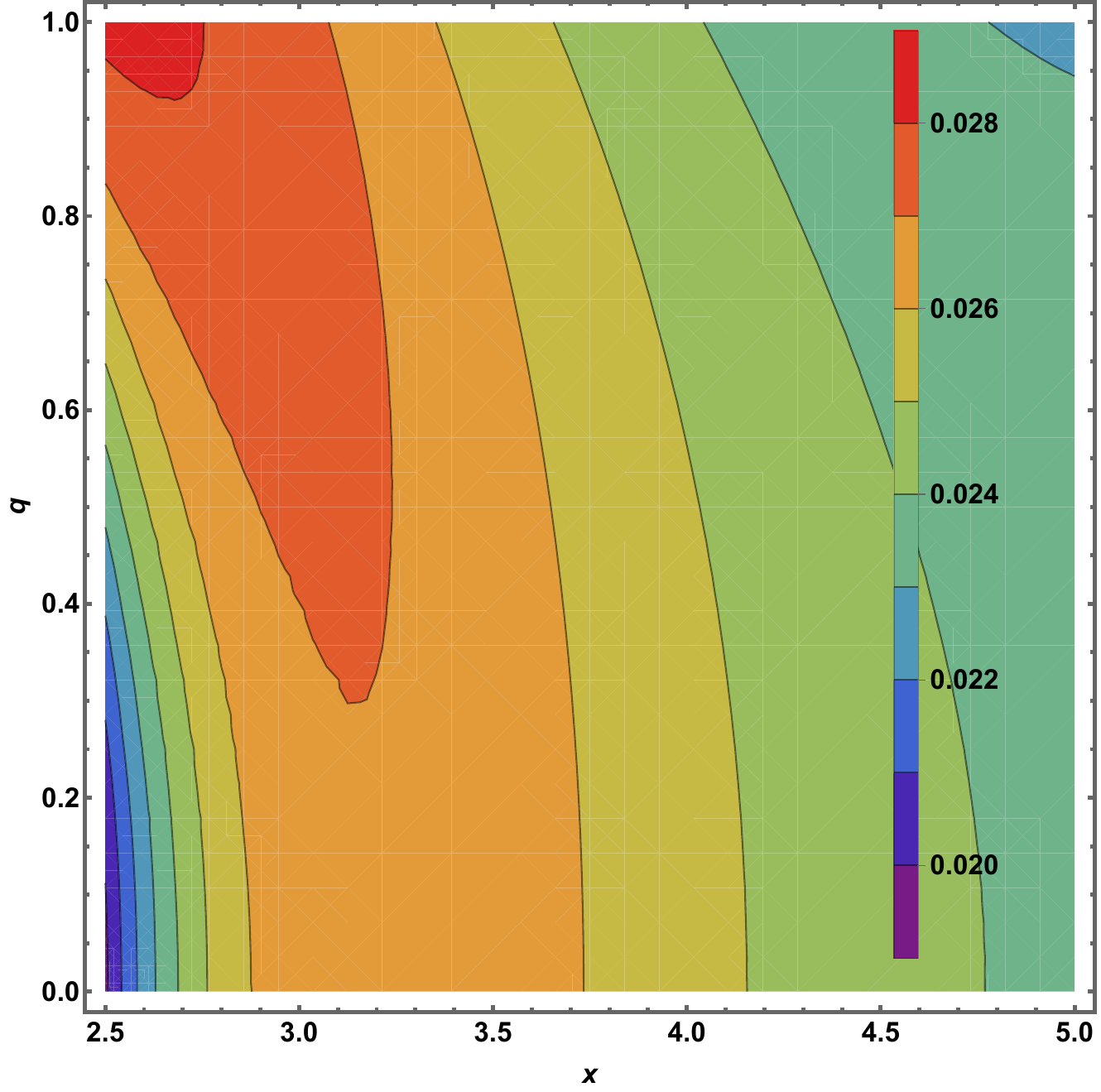}\qquad
    \includegraphics[width=0.4\linewidth]{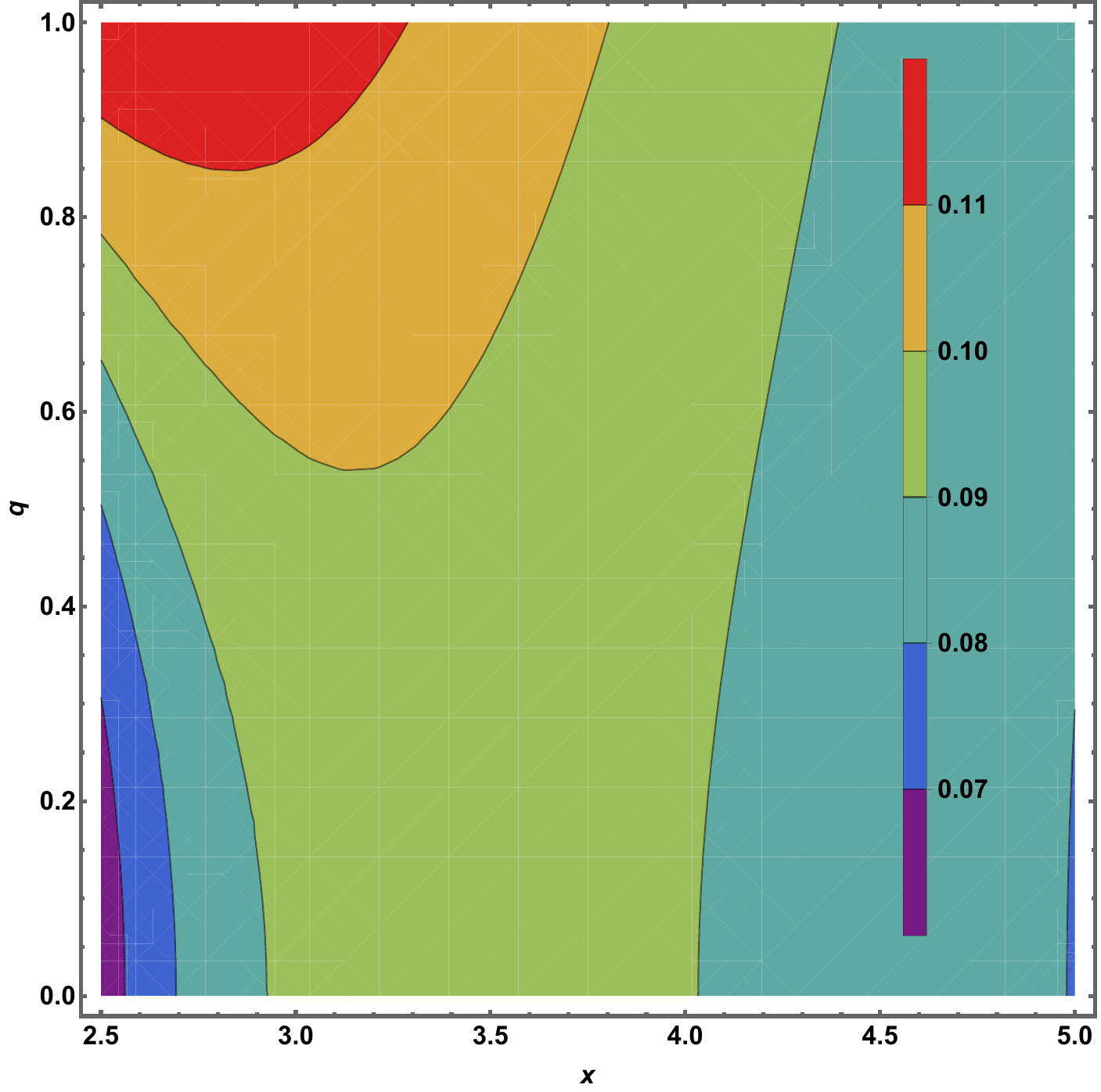}\\
    (i) $\ell=0$ \hspace{8cm} (ii) $\ell=1$
    \caption{\footnotesize Contour plot of $M^2\,\mathcal{V}_\text{scalar}$ for $\alpha=0.10$. Here $\eta=0.1,\,\zeta=0.01$. Comparing with Fig.~\ref{fig:5}, the increased $\alpha$ value shifts the contour pattern toward lower potential values, consistent with the weakening of effective gravity due to the CS.}
    \label{fig:6}
\end{figure}

To elucidate the parameter dependence of the scalar potential, we present a comprehensive graphical analysis in Figs.~\ref{fig:1}--\ref{fig:6}. Figures~\ref{fig:1} and \ref{fig:2} display the behavior of $\mathcal{V}_\text{scalar}$ as a function of radial coordinate $r$ for the $\ell=0$ (monopole) and $\ell=1$ (dipole) modes, respectively, under variations of the CS parameter $\alpha$, quintessence equation-of-state parameter $w$, and NLED charge parameter $k$. For the monopole mode (Fig.~\ref{fig:1}), we observe that increasing $\alpha$ generally reduces the potential barrier height, reflecting the weakening of effective gravitational attraction due to the CS. Conversely, varying $w$ toward more negative values (stronger quintessence effects) modifies the potential shape at intermediate radii. The NLED parameter $k$ introduces subtle modifications to the potential peak position and width. For the dipole mode (Fig.~\ref{fig:2}), the centrifugal barrier becomes prominent, and similar parameter dependencies are observed, though with enhanced potential magnitudes due to the $\ell(\ell+1)/r^2$ term. Figure~\ref{fig:3} provides an additional perspective on the $\ell=2$ potential under different parameter combinations, confirming the systematic trends.

To provide a more global view of the potential structure, Fig.~\ref{fig:4} presents three-dimensional surface plots of the dimensionless potential $M^2\mathcal{V}_\text{scalar}$ as a function of the scaled radial coordinate $x = r/M$ and the charge-to-mass ratio $q = Q/M$ for both monopole ($\ell=0$) and dipole ($\ell=1$) modes. These plots reveal how the potential landscape varies across the $(x, q)$ parameter space for different values of the CS parameter $\alpha$ (yellow: $\alpha=0.05$, blue: $\alpha=0.1$, green: $\alpha=0.15$). The layered structure clearly demonstrates that increasing $\alpha$ systematically lowers the potential surface, consistent with the reduced effective gravitational strength induced by the CS. Figures~\ref{fig:5} and \ref{fig:6} complement this analysis with contour plots of $M^2\mathcal{V}_\text{scalar}$ for fixed $\alpha=0.05$ and $\alpha=0.10$, respectively, providing a detailed view of the potential gradient structure in the $(x, q)$ plane. These visualizations facilitate the identification of regions in parameter space where the potential exhibits suitable barrier structures for GBF analyses.

For the specific quintessence equation-of-state parameter $w=-2/3$, a physically motivated value corresponding to network of cosmic strings or frustrated domain walls, the perturbative potential (\ref{ff5}) simplifies to:
\begin{eqnarray}
\mathcal{V}_\text{scalar}=\left[\frac{\ell\,(\ell+1)}{r^2}+\frac{2\,M\,(r-k)}{r^4}\,e^{-k/r}-\frac{\mathrm{N}}{r}-\frac{2\,\Lambda}{3}\right]\,\left(1-\alpha-\frac{2\,M}{r}\,e^{-k/r}-\mathrm{N}\,r-\frac{\Lambda}{3}\,r^2\right).\label{ff6}
\end{eqnarray}
Introducing dimensionless variables $x=r/M$ (scaled radial coordinate), $q=Q/M$ (charge-to-mass ratio), $\eta=M\sqrt{-\Lambda/3}$ (scaled cosmological constant), and $\zeta=M\,\mathrm{N}$ (scaled quintessence parameter), we can express Eq.~(\ref{ff6}) in the compact dimensionless form:
\begin{align}
    M^2\,\mathcal{V}_\text{scalar}=\left[\frac{\ell\,(\ell+1)}{x^2}+\frac{(2\,x-q^2)}{x^4}\,e^{-\frac{q^2}{2\,x}}-\frac{\zeta}{x}+2\,\eta^2\right]\,\left(1-\alpha-\frac{2}{x}\,e^{-\frac{q^2}{2\,x}}-\zeta\,x+\eta^2\,x^2\right).\label{ff7}
\end{align}
This dimensionless formulation facilitates numerical computations and provides a universal framework for analyzing the potential structure across different mass scales.

\subsection{Electromagnetic Field Dynamics: Spin-1 Perturbations}

We now turn to the analysis of electromagnetic perturbations in the AdS-NLED-CS-QF BH background. The electromagnetic field in curved spacetime is governed by Maxwell's equations generalized to arbitrary geometries \cite{ref69}:
\begin{equation}
    \frac{1}{\sqrt{-g}}\,\partial_{\mu}\left(\sqrt{-g}\,g^{\mu\sigma}\,g^{\nu\tau}\,F_{\sigma\tau}\right)=0,\label{em1}
\end{equation}
where $F_{\sigma\tau}=\partial_{\sigma}A_{\tau}-\partial_{\tau}A_{\sigma}$ is the electromagnetic field tensor and $A_{\mu}$ is the four-vector potential. To analyze perturbations in a spherically symmetric background, we employ the Regge-Wheeler-Zerilli formalism, decomposing $A_{\mu}$ into scalar and vector spherical harmonics:
\begin{equation}
    A_{\mu}=\sum_{\ell,m}\,e^{-i\,\omega\,t}\,\left(\left[\begin{array}{c}
         0\\
         0\\
         \psi_{em}(r)\,{\bf S}_{\ell,m}
    \end{array}\right]+\left[\begin{array}{c}
         j^{\ell,m}(r)\,Y^{m}_{\ell}\\
         h^{\ell,m}(r)\,Y^{m}_{\ell}\\
         k^{\ell,m}(r)\,{\bf Y}^{m}_{\ell}\\
    \end{array}\right] 
    \right),\label{em2}
\end{equation}
where $Y^{m}_{\ell}$ are scalar spherical harmonics and $({\bf S}_{\ell,m}, {\bf Y}_{\ell,m})$ are vector harmonics defined by:
\begin{equation}
    {\bf S}_{\ell,m}=\left(\begin{array}{c}
         \frac{1}{\sin \theta}\,\partial_{\phi}\,Y^{m}_{\ell}\\
         -\sin \theta\,\partial_{\theta}\,Y^{m}_{\ell} 
    \end{array}\right),\quad\quad {\bf Y}_{\ell,m}=\left(\begin{array}{c}
         \partial_{\theta}\,Y^{m}_{\ell}\\
         \partial_{\phi}\,Y^{m}_{\ell} 
    \end{array}\right).\label{em3}
\end{equation}

The first term in Eq.~(\ref{em2}) corresponds to axial (odd-parity) perturbations with parity $(-1)^{\ell+1}$, while the second term represents polar (even-parity) perturbations with parity $(-1)^{\ell}$. A fundamental result in BH perturbation theory is that axial and polar modes contribute equally to physical observables \cite{ref70,ref71}, allowing us to focus exclusively on the axial sector without loss of generality. Substituting the axial mode decomposition into Eq.~(\ref{em1}) and transforming to the tortoise coordinate $r_*$, we obtain the Schrödinger-like radial equation:
\begin{equation}
    \frac{\partial^2 \psi_{em}(r_*)}{\partial r^2_{*}}+\left(\omega^2-\mathcal{V}_{em}\right)\,\psi_{em}(r_*)=0,\label{em4}
\end{equation}
where the electromagnetic effective potential is given by:
\begin{equation}
    \mathcal{V}^{\ell}_{em}=\frac{\ell\,(\ell+1)}{r^2}\,f(r)=\frac{\ell\,(\ell+1)}{r^2}\,\left(1-\alpha-\frac{2\,M}{r}\,e^{-k/r}-\frac{\mathrm{N}}{r^{3\,w+1}}-\frac{\Lambda}{3}\,r^2\right).\label{em5}
\end{equation}

Comparing Eq.~(\ref{em5}) with the scalar potential in Eq.~(\ref{ff5}), we observe that the electromagnetic potential has a simpler structure, lacking the $f'(r)/r$ term present in the scalar case. This structural difference arises from the vector nature of the electromagnetic field and reflects the distinct coupling between spin-1 fields and spacetime curvature. Nevertheless, the electromagnetic potential remains sensitive to all the same physical parameters: NLED charge $k$, CS parameter $\alpha$, quintessence parameters $(\mathrm{N}, w)$, BH mass $M$, cosmological constant $\Lambda$, and angular momentum quantum number $\ell$.

\begin{figure}[ht!]
    \centering
    \includegraphics[width=0.325\linewidth]{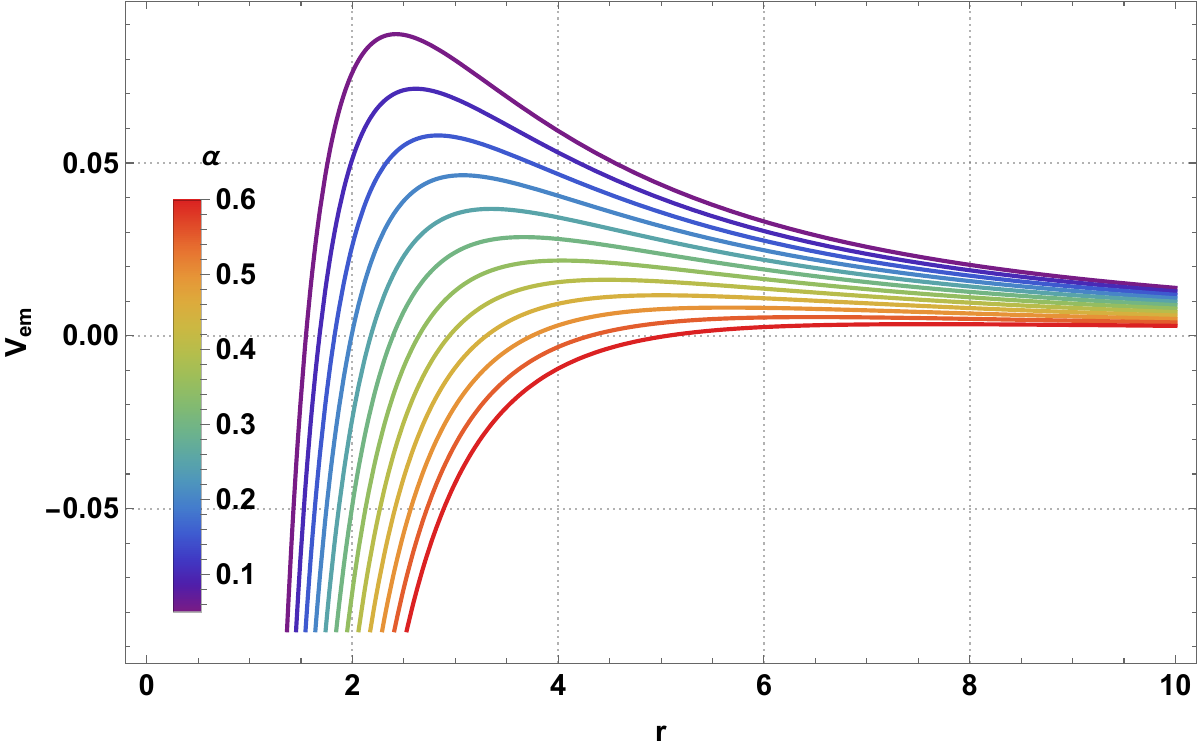}
    \includegraphics[width=0.325\linewidth]{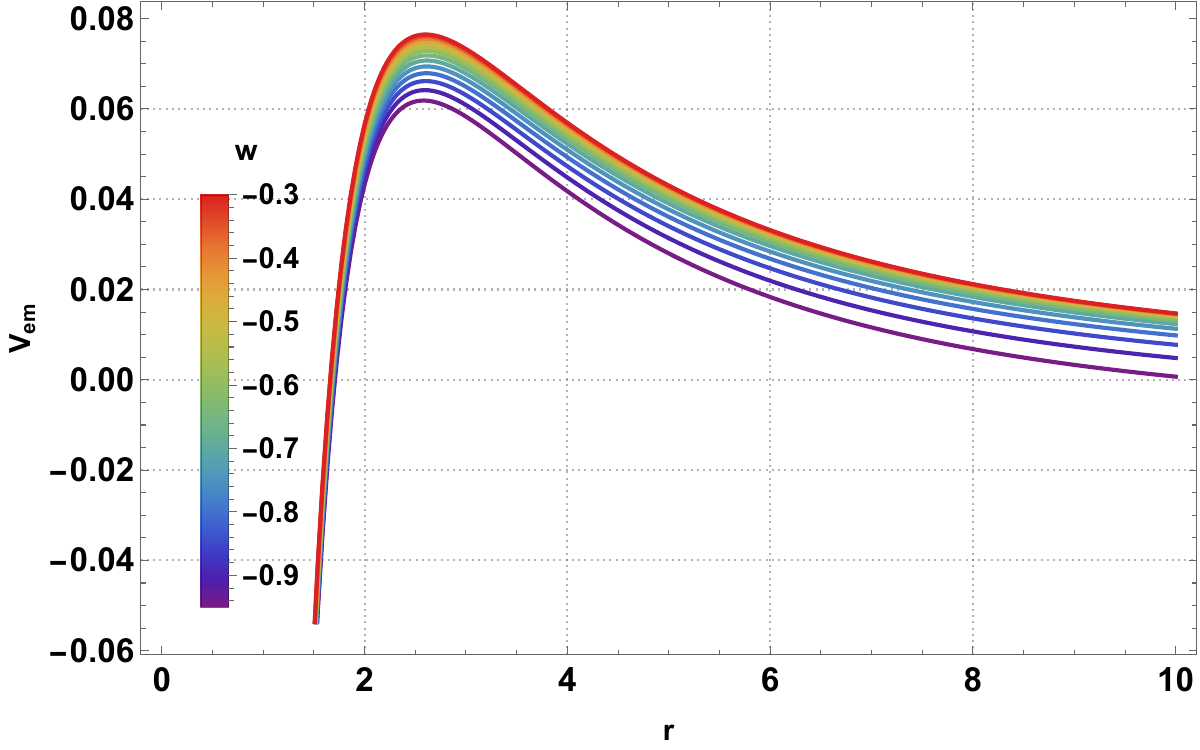}
    \includegraphics[width=0.325\linewidth]{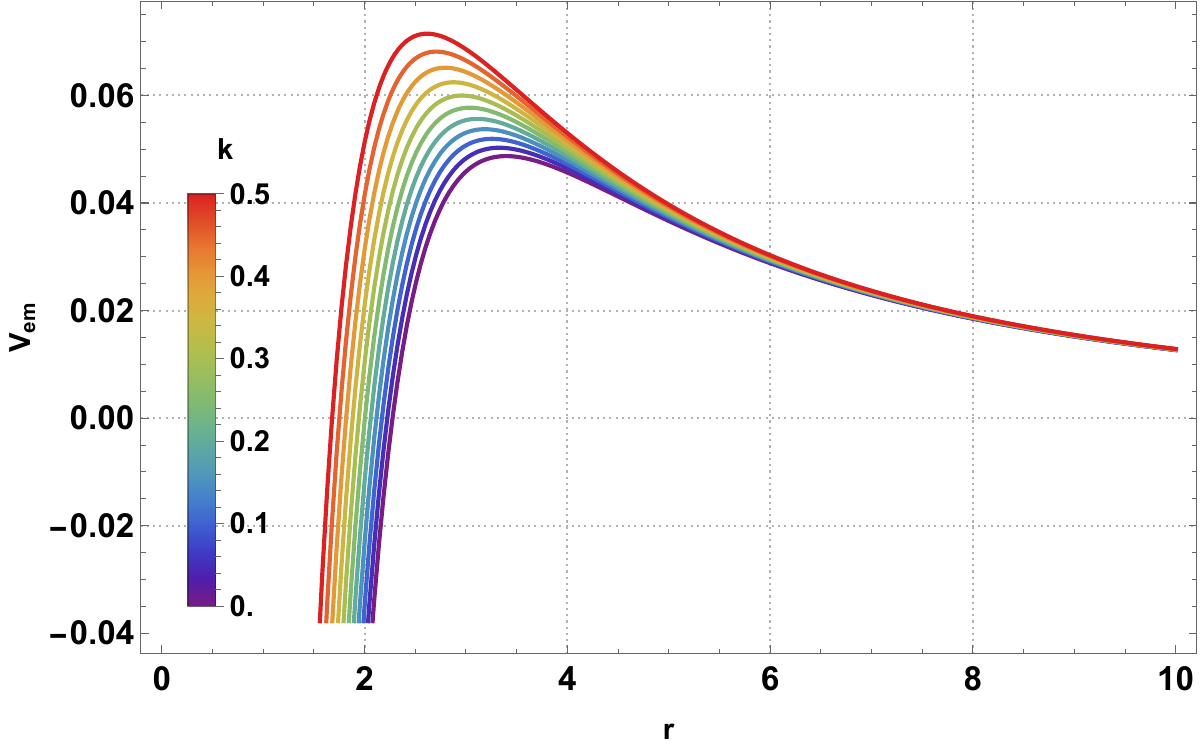}\\
    (i) $Q=1,\,w=-2/3$ \hspace{3cm} (ii) $\alpha=0.1,\,k=0.5$ \hspace{3cm} (iii) $\alpha=0.1,\,w=-2/3$
    \caption{\footnotesize Behavior of the electromagnetic perturbative potential $\mathcal{V}_\text{em}$ for $\ell=1$-state by varying values of $\alpha, w$ and $k$. Here $M=1,\,\Lambda=-0.001,\,\mathrm{N}=0.01$. The electromagnetic potential exhibits qualitatively similar parameter dependencies as the scalar case but with a simpler functional form due to the absence of the $f'/r$ term.}
    \label{fig:7}
\end{figure}

\begin{figure}[ht!]
    \centering
    \includegraphics[width=0.325\linewidth]{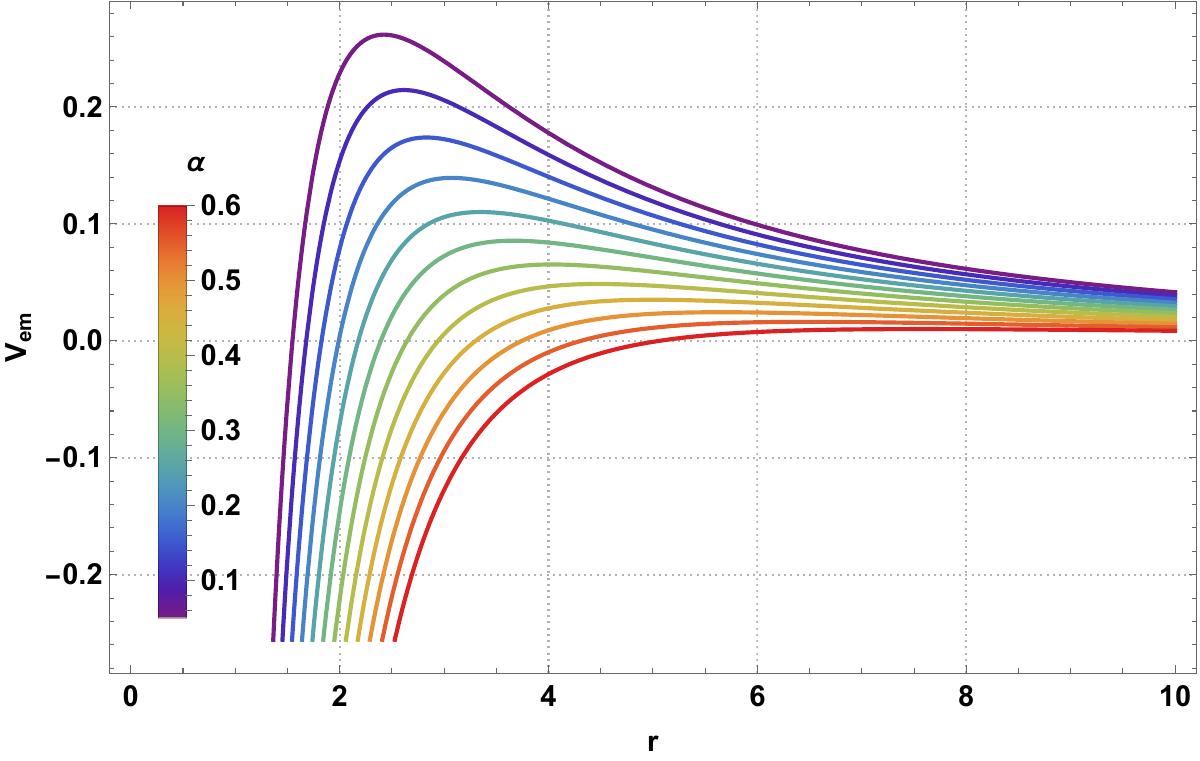}
    \includegraphics[width=0.325\linewidth]{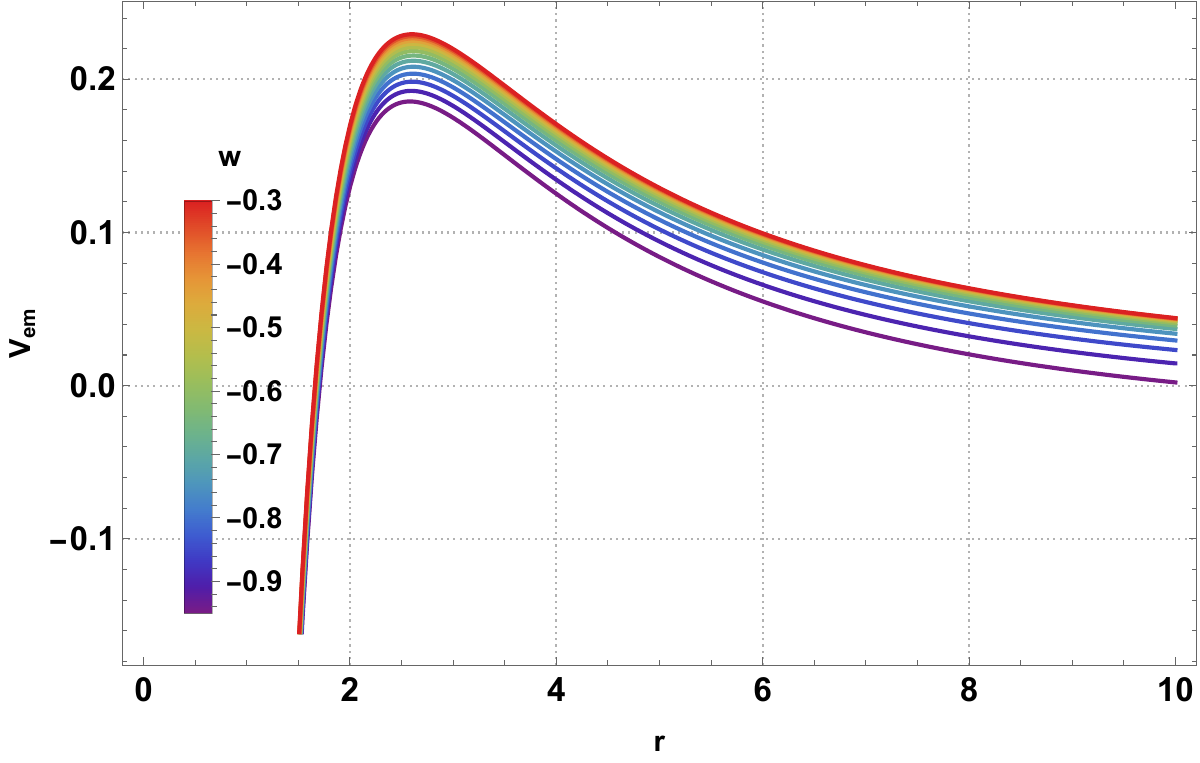}
    \includegraphics[width=0.325\linewidth]{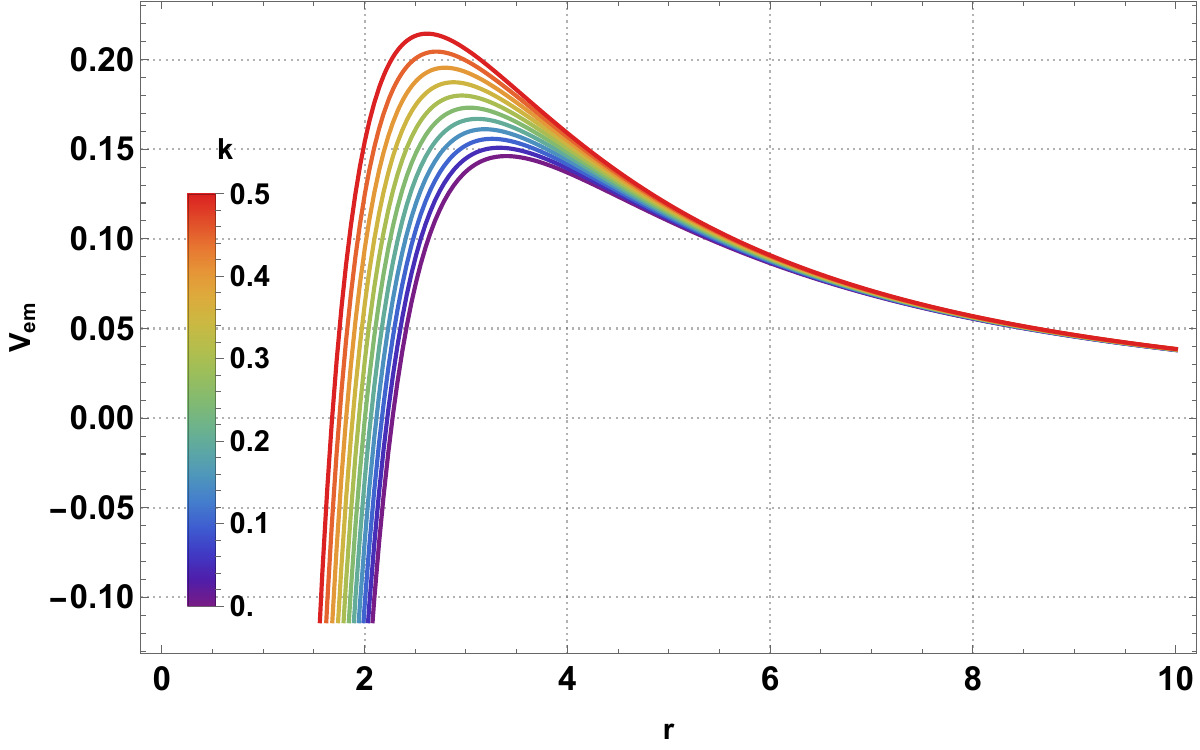}\\
    (i) $Q=1,\,w=-2/3$ \hspace{3cm} (ii) $\alpha=0.1,\,k=0.5$ \hspace{3cm} (iii) $\alpha=0.1,\,w=-2/3$
    \caption{\footnotesize Behavior of the electromagnetic perturbative potential $\mathcal{V}_\text{em}$ for $\ell=2$-state by varying values of $\alpha, w$ and $k$. The quadrupole mode exhibits enhanced potential barriers compared to $\ell=1$ due to the stronger centrifugal term.}
    \label{fig:8}
\end{figure}

\begin{figure}[ht!]
    \centering
    \includegraphics[width=0.4\linewidth]{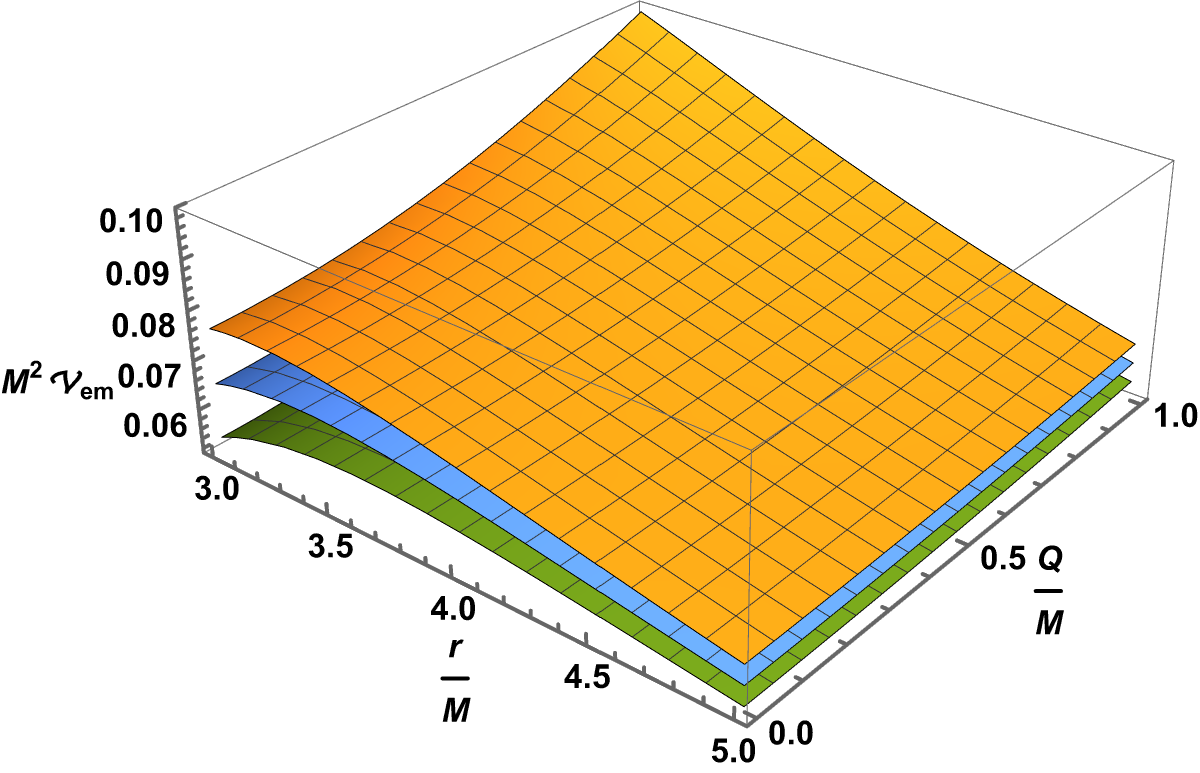}\qquad
    \includegraphics[width=0.4\linewidth]{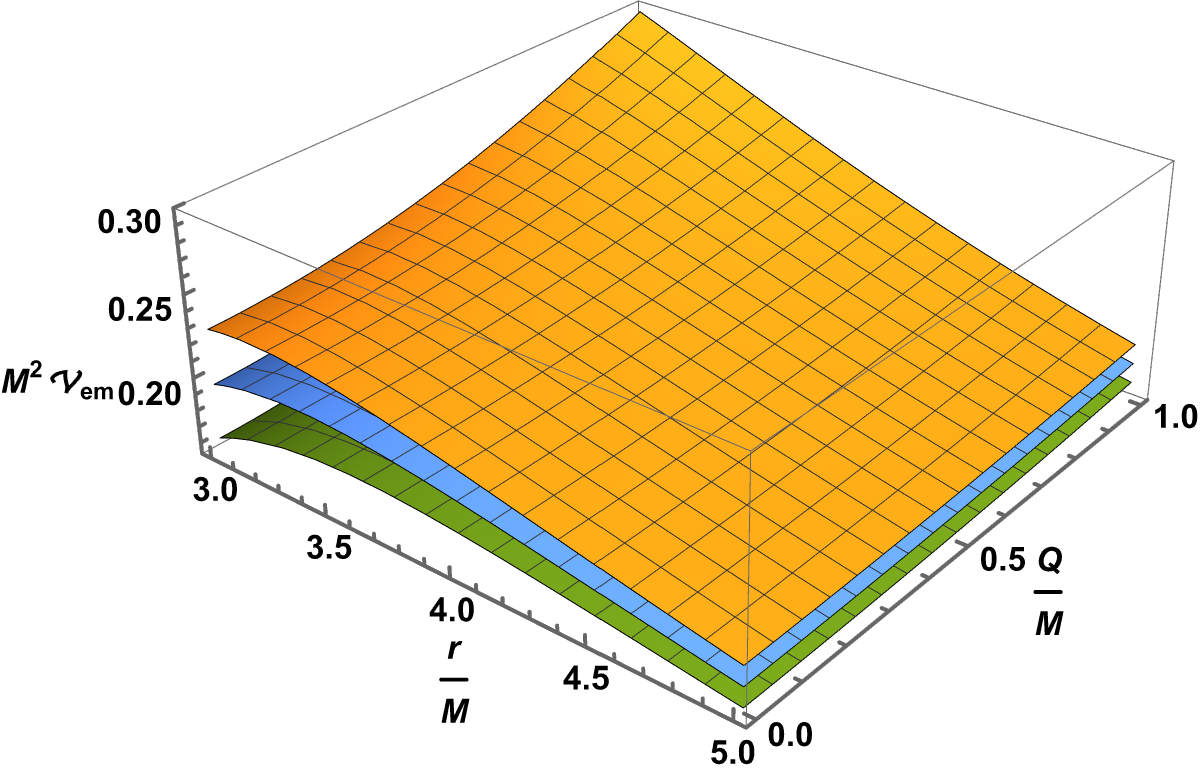}\\
    (i) $\ell=1$ \hspace{8cm} (ii) $\ell=2$
    \caption{\footnotesize Three-dimensional plot of $M^2\,\mathcal{V}_\text{em}$ as a function of $(r/M, Q/M)$. Here $\eta=0.1,\,\zeta=0.01$. Yellow $\to \alpha=0.05$; Blue $\to \alpha=0.1$; Green $\to \alpha=0.15$. The electromagnetic potential surfaces exhibit the same qualitative CS-induced lowering as observed in the scalar case (Fig.~\ref{fig:4}).}
    \label{fig:9}
\end{figure}

\begin{figure}[ht!]
    \centering
    \includegraphics[width=0.4\linewidth]{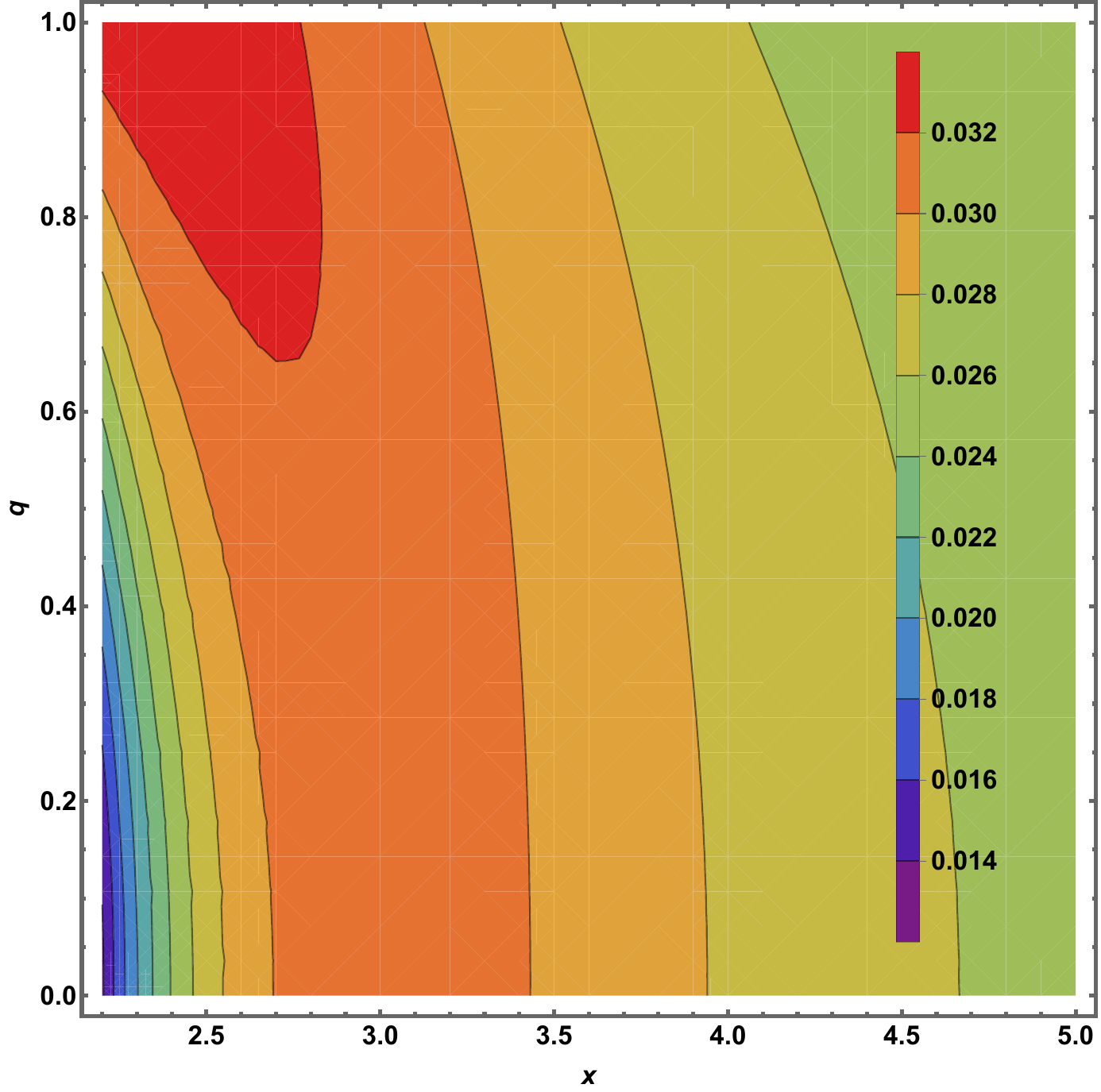}\qquad
    \includegraphics[width=0.4\linewidth]{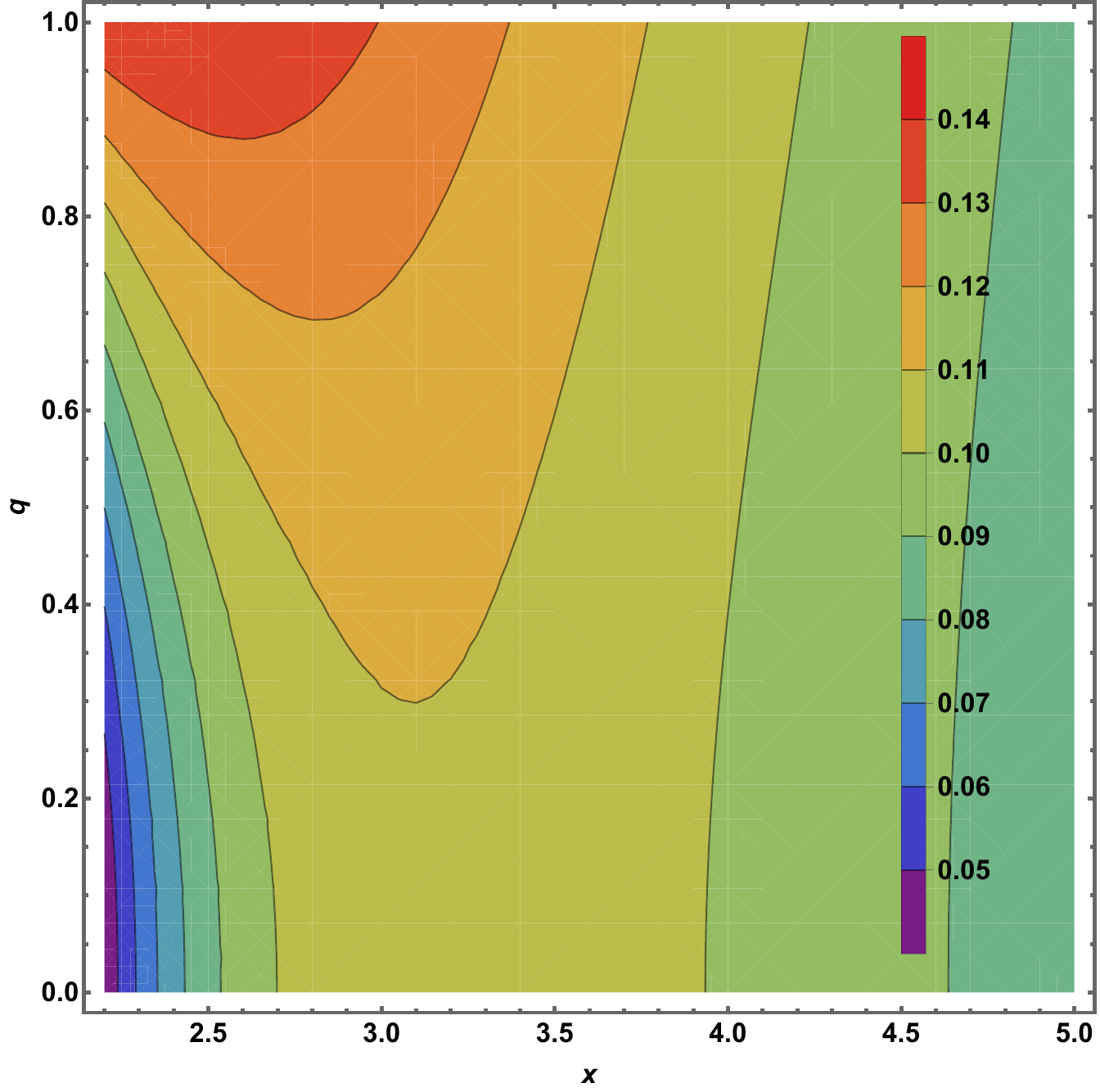}\\
    (i) $\alpha=0.05$ \hspace{8cm} (ii) $\alpha=0.1$
    \caption{\footnotesize Contour plot of $M^2\,\mathcal{V}_\text{em}$ for $\ell=1$. Here $\eta=0.1,\,\zeta=0.01$. The contour patterns reveal how the electromagnetic potential landscape shifts systematically with increasing CS parameter $\alpha$.}
    \label{fig:10}
\end{figure}

Figures~\ref{fig:7} and \ref{fig:8} illustrate the behavior of $\mathcal{V}_{em}$ for the $\ell=1$ and $\ell=2$ modes, respectively, under systematic variations of $\alpha$, $w$, and $k$. The trends observed mirror those for scalar perturbations: increasing $\alpha$ lowers the potential barrier, more negative $w$ values modify the intermediate-radius behavior, and larger $k$ shifts the potential peak structure. Figure~\ref{fig:9} provides a three-dimensional visualization of $M^2\mathcal{V}_{em}$ in the $(r/M, Q/M)$ plane for different $\alpha$ values, confirming the systematic reduction of the potential surface with increasing CS parameter. Figure~\ref{fig:10} presents contour plots for $\ell=1$ at $\alpha=0.05$ and $\alpha=0.1$, facilitating detailed examination of the potential gradient structure.

For the specific case $w=-2/3$, the electromagnetic potential simplifies to:
\begin{eqnarray}
\mathcal{V}_{em}=\frac{\ell\,(\ell+1)}{r^2}\,\left(1-\alpha-\frac{2\,M}{r}\,e^{-k/r}-\mathrm{N}\,r-\frac{\Lambda}{3}\,r^2\right),\label{em6}
\end{eqnarray}
which in dimensionless variables becomes:
\begin{align}
    M^2\,\mathcal{V}_\text{em}=\frac{\ell\,(\ell+1)}{x^2}\,\left(1-\alpha-\frac{2}{x}\,e^{-\frac{q^2}{2\,x}}-\zeta\,x+\eta^2\,x^2\right).\label{em7}
\end{align}

\subsection{Fermionic Field Dynamics: Spin-1/2 Perturbations}

Finally, we analyze spin-1/2 fermionic perturbations governed by the Dirac equation in the AdS-NLED-CS-QF BH background. Fermionic field perturbations have been investigated in various BH spacetimes, including Schwarzschild, Reissner-Nordström, and rotating Kerr geometries, providing valuable insights into the coupling between fermion spin and spacetime curvature \cite{ref72,ref73}. The massless Dirac equation in curved spacetime is given by:
\begin{equation}
\left[i\,\gamma^{a}\,e^{\mu}_{a}\,(\partial_{\mu}+\Gamma_{\mu})-\mu\right]\,\Psi=0,\label{fermi1}
\end{equation}
where $\mu$ is the Dirac field mass (which we set to zero for massless fermions), $e^{\mu}_{a}$ is the inverse tetrad satisfying $g_{\mu\nu}=\eta_{ab}\,e^{a}_{\mu}\,e^{b}_{\nu}$ (with $\eta_{ab}$ being the Minkowski metric), $\gamma^{a}$ are the Dirac gamma matrices, and $\Gamma_{\mu}$ is the spin connection encoding the coupling between the fermion spin and spacetime torsion.

For the spherically symmetric BH metric in Eq.~(\ref{aa1}), the Dirac equation can be reduced to a one-dimensional Schrödinger-like wave equation through standard techniques involving separation of variables and spinor decomposition \cite{ref72}:
\begin{equation}
\frac{d^{2}U}{dr_{\ast }^{2}}+\left( \omega^{2}-\mathcal{V}_{\pm 1/2}\right)\, U=0,  \label{fermi2}
\end{equation}
where the effective potential for massless spin-1/2 perturbations is:
\begin{equation}
\mathcal{V}_{\pm 1/2}=\frac{\left( \frac{1}{2}+\ell\right) }{r^{2}}\,\left[\ell+\frac{1}{2} \pm \frac{r\,f'(r)}{2\,\sqrt{f(r)}}\mp \sqrt{f(r)}\right]\,f(r).\label{fermi3}
\end{equation}
The $\pm$ subscript in $\mathcal{V}_{\pm 1/2}$ denotes the two possible helicity states of the fermion: positive helicity (spin aligned with momentum) and negative helicity (spin anti-aligned with momentum). This helicity dependence, absent in bosonic (scalar and electromagnetic) perturbations, arises from the intrinsic spin-1/2 nature of fermions and leads to distinct effective potentials for the two helicity channels.

\begin{figure}[ht!]
    \centering
    \includegraphics[width=0.325\linewidth]{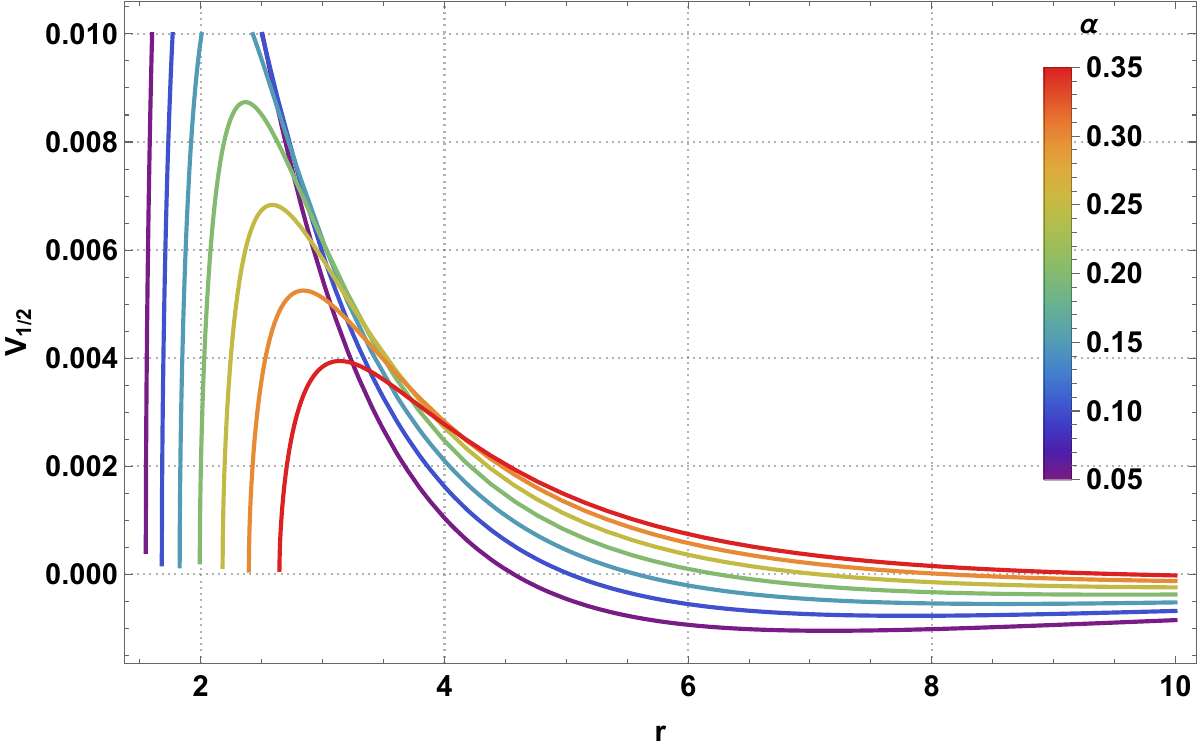}
    \includegraphics[width=0.325\linewidth]{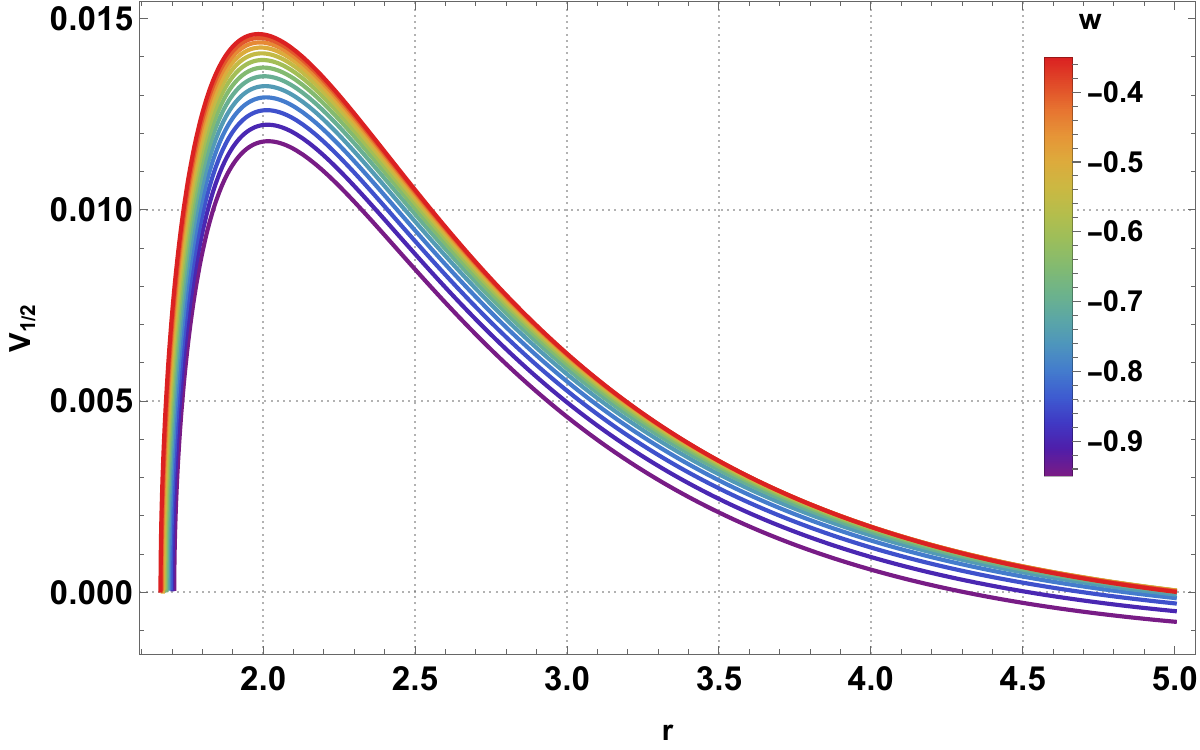}
    \includegraphics[width=0.325\linewidth]{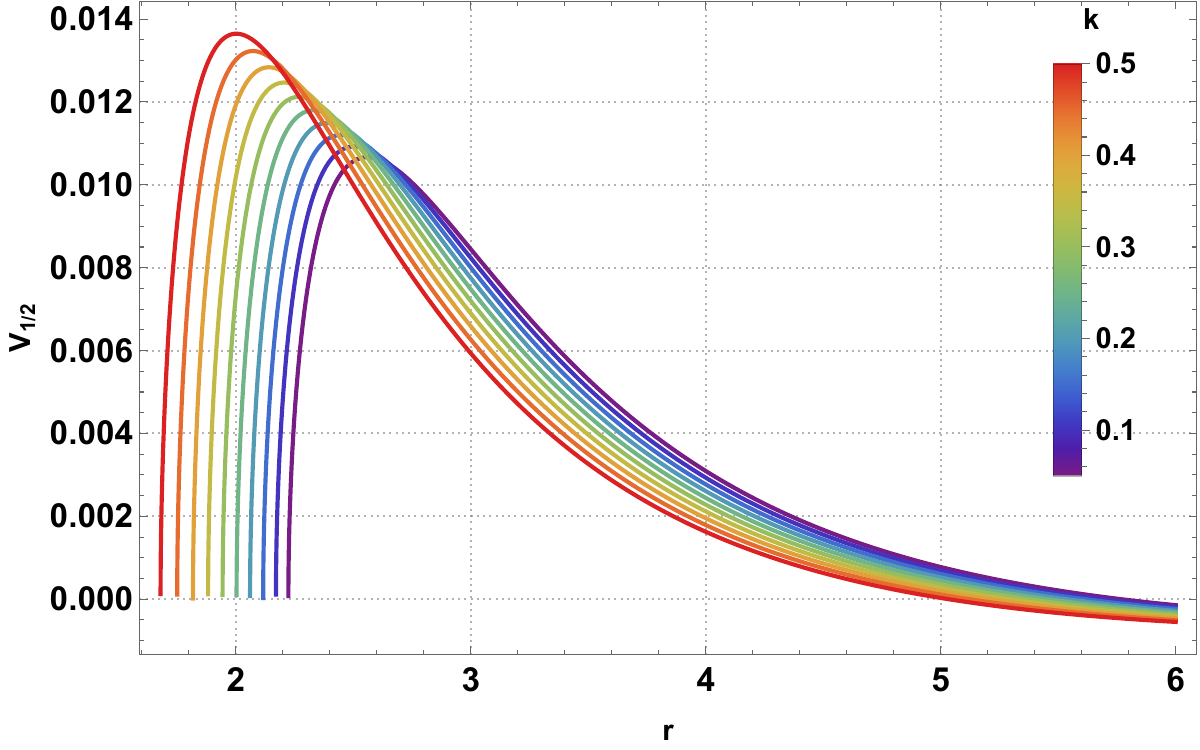}\\
    (i) $Q=1,\,w=-2/3$ \hspace{3cm} (ii) $\alpha=0.1,\,k=0.5$ \hspace{3cm} (iii) $\alpha=0.1,\,w=-2/3$
    \caption{\footnotesize Behavior of the positive helicity fermionic field perturbative potential $\mathcal{V}_{+1/2}$ for $\ell=0$-state by varying values of $\alpha, w$ and $k$. Here $M=1,\,\Lambda=-0.001,\,\mathrm{N}=0.01$. The fermionic potential exhibits distinct characteristics due to spin-curvature coupling terms absent in scalar and electromagnetic cases.}
    \label{fig:11}
\end{figure}

\begin{figure}[ht!]
    \centering
    \includegraphics[width=0.325\linewidth]{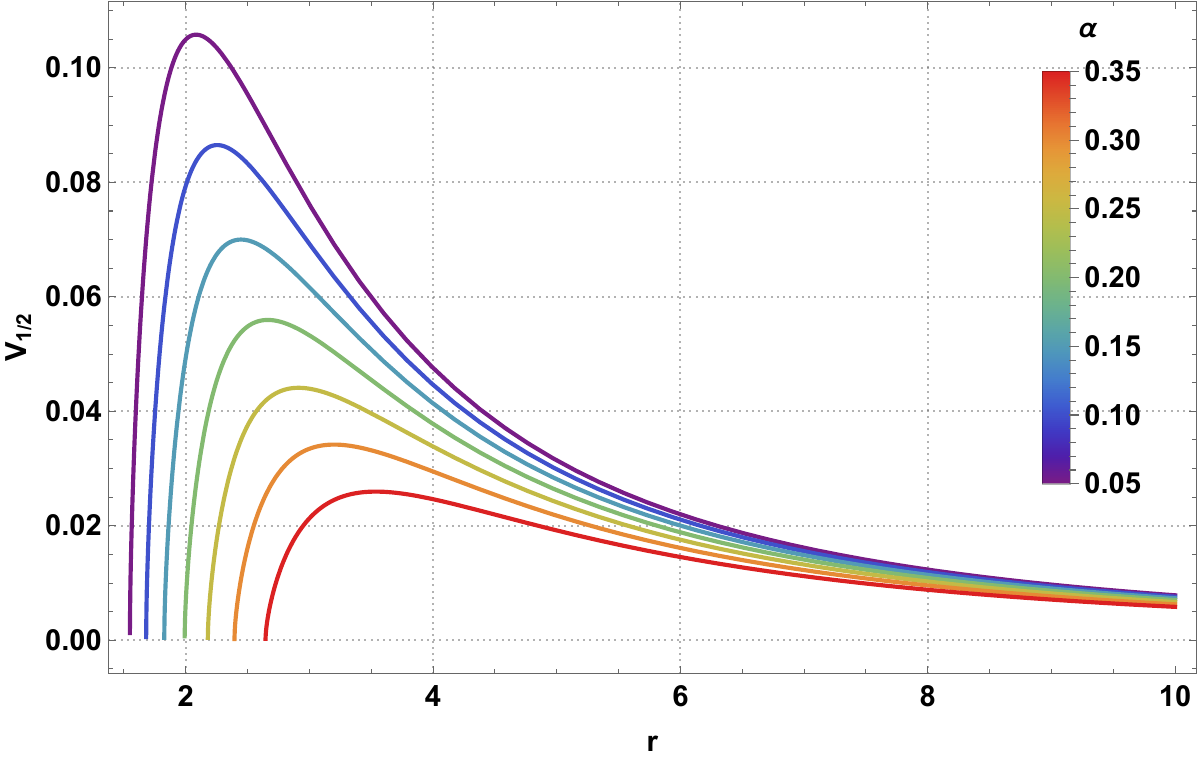}
    \includegraphics[width=0.325\linewidth]{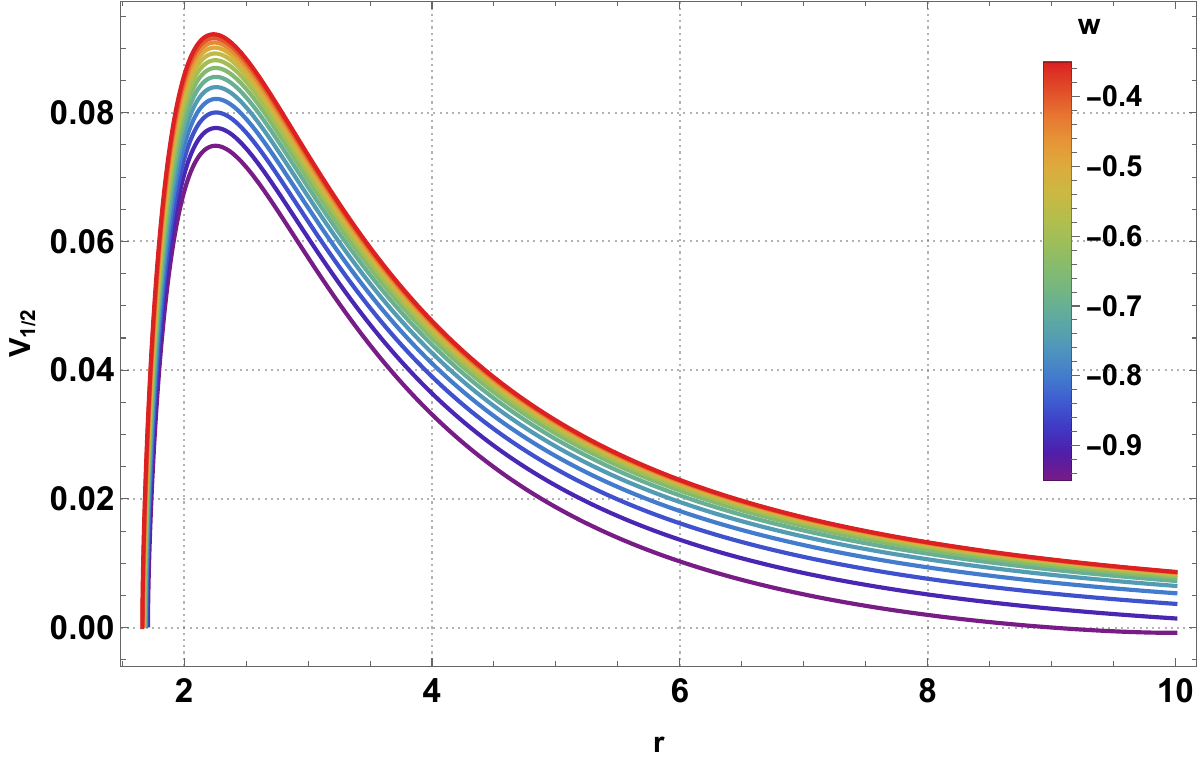}
    \includegraphics[width=0.325\linewidth]{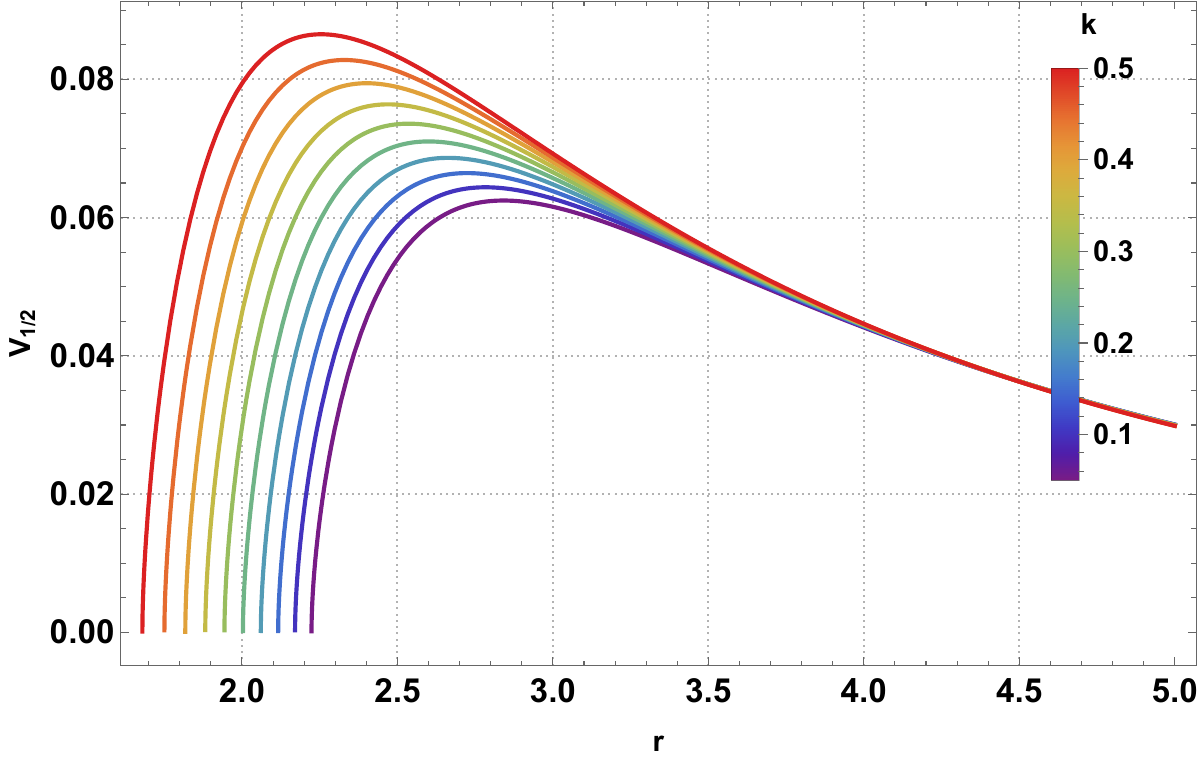}\\
    (i) $Q=1,\,w=-2/3$ \hspace{3cm} (ii) $\alpha=0.1,\,k=0.5$ \hspace{3cm} (iii) $\alpha=0.1,\,w=-2/3$
    \caption{\footnotesize Behavior of the positive helicity fermionic field perturbative potential $\mathcal{V}_{+1/2}$ for $\ell=1$-state by varying values of $\alpha, w$ and $k$. The dipole mode shows enhanced complexity in the potential structure compared to the monopole case in Fig.~\ref{fig:11}.}
    \label{fig:12}
\end{figure}

\begin{figure}[ht!]
    \centering
    \includegraphics[width=0.325\linewidth]{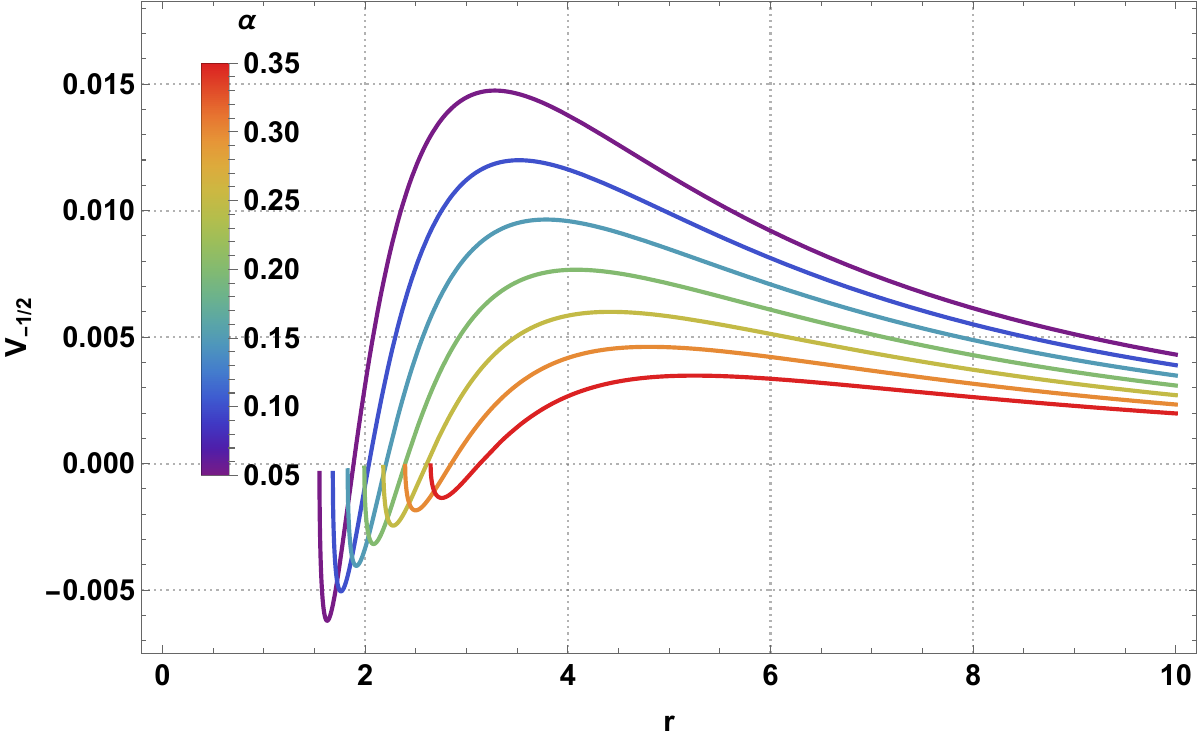}
    \includegraphics[width=0.325\linewidth]{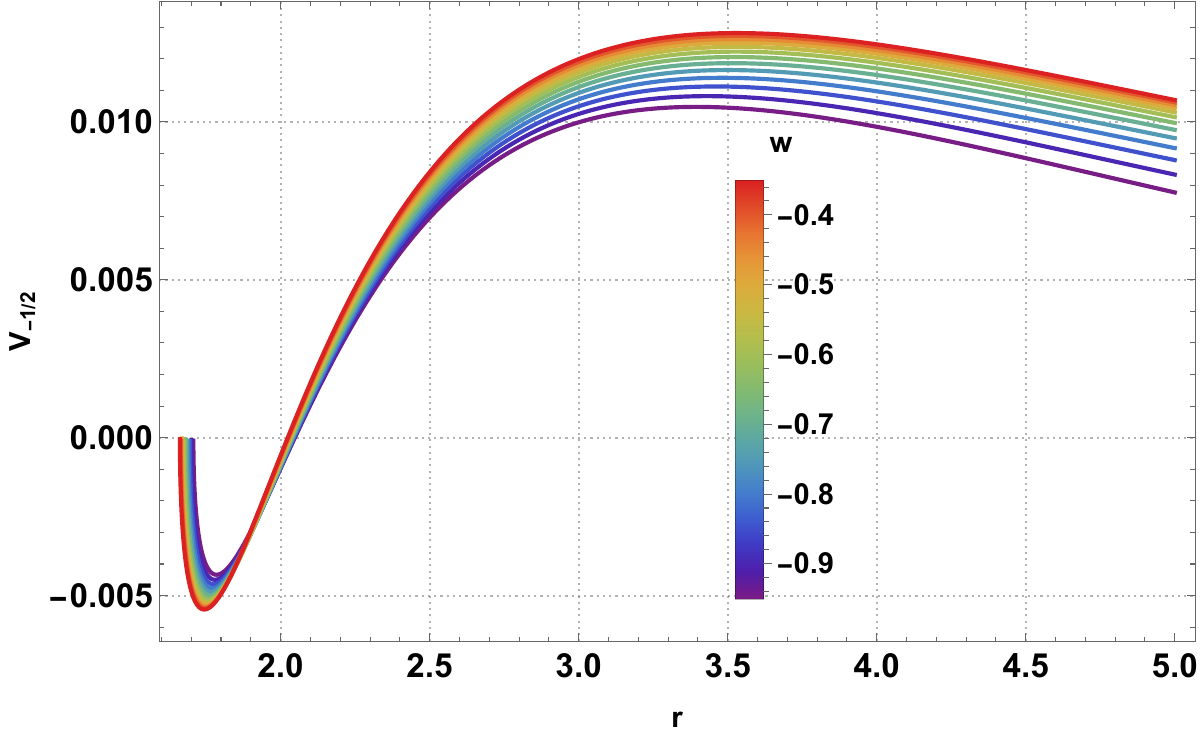}
    \includegraphics[width=0.325\linewidth]{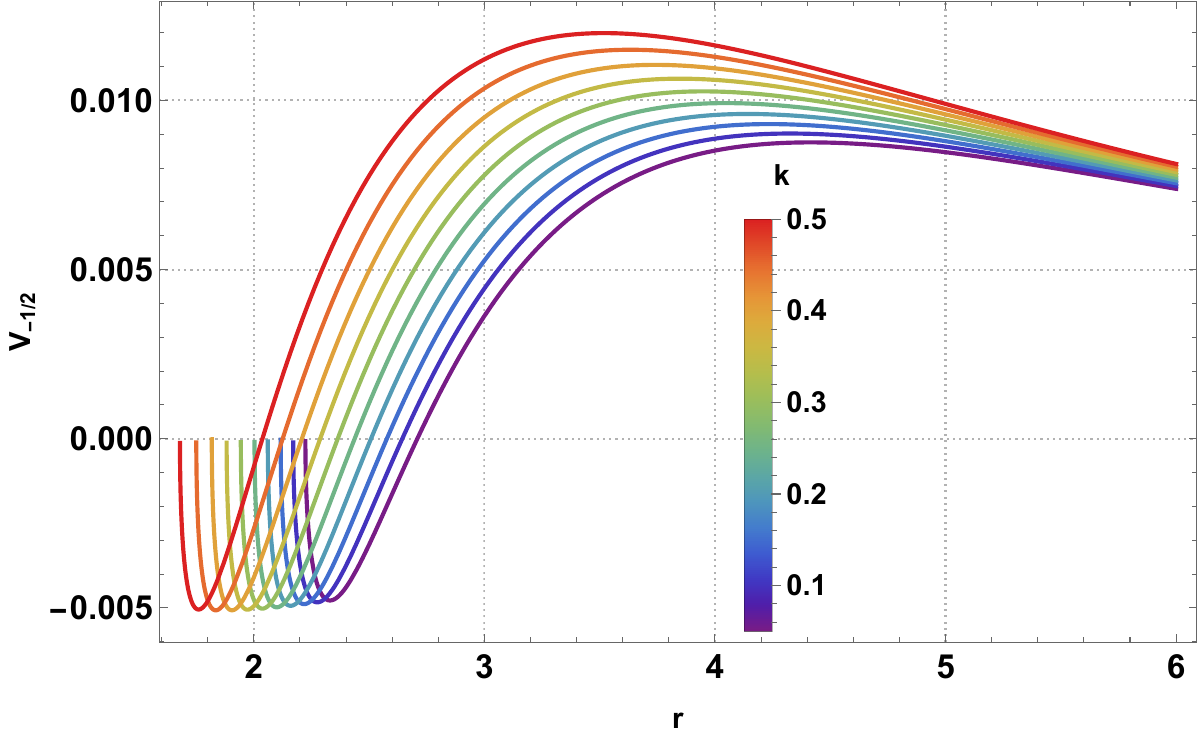}\\
    (i) $Q=1,\,w=-2/3$ \hspace{3cm} (ii) $\alpha=0.1,\,k=0.5$ \hspace{3cm} (iii) $\alpha=0.1,\,w=-2/3$
    \caption{\footnotesize Behavior of the negative helicity fermionic field perturbative potential $\mathcal{V}_{-1/2}$ for $\ell=0$-state by varying values of $\alpha, w$ and $k$. Here $M=1,\,\Lambda=-0.001,\,\mathrm{N}=0.01$. Comparing with Fig.~\ref{fig:11}, the negative helicity potential exhibits distinct features due to the opposite sign in the spin-curvature coupling terms in Eq.~(\ref{fermi3}).}
    \label{fig:13}
\end{figure}

\begin{figure}[ht!]
    \centering
    \includegraphics[width=0.325\linewidth]{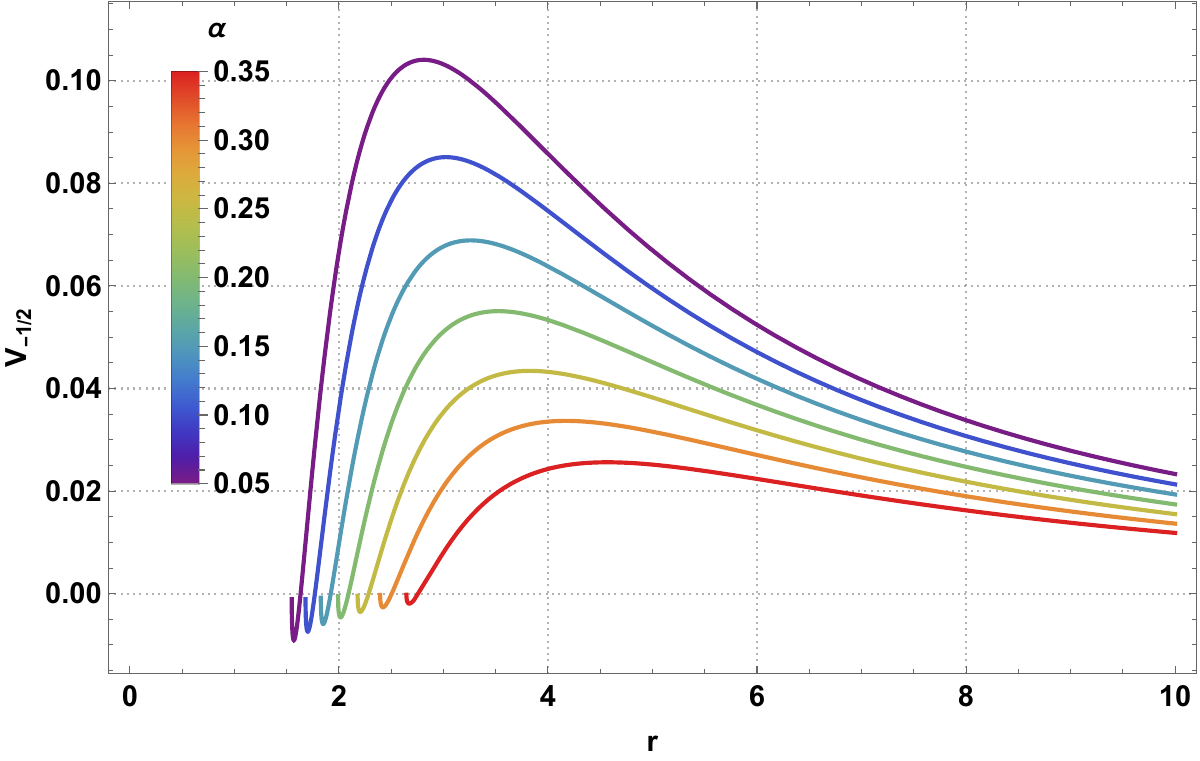}
    \includegraphics[width=0.325\linewidth]{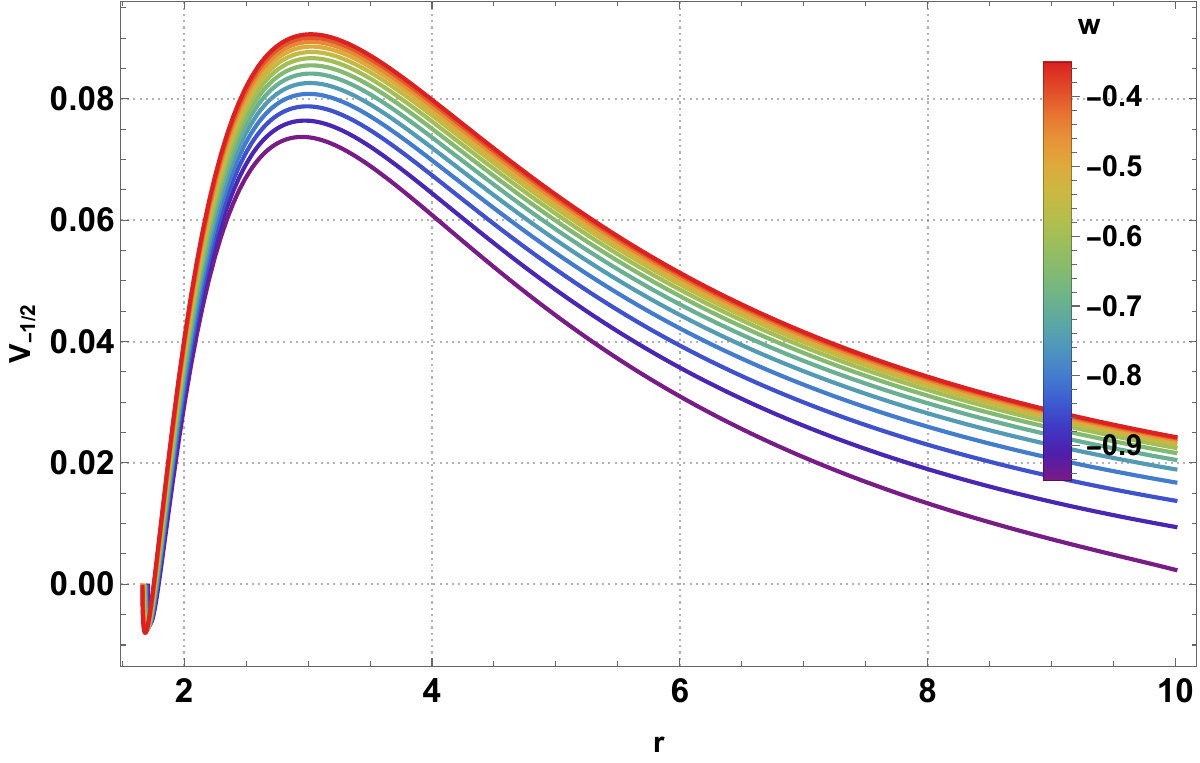}
    \includegraphics[width=0.325\linewidth]{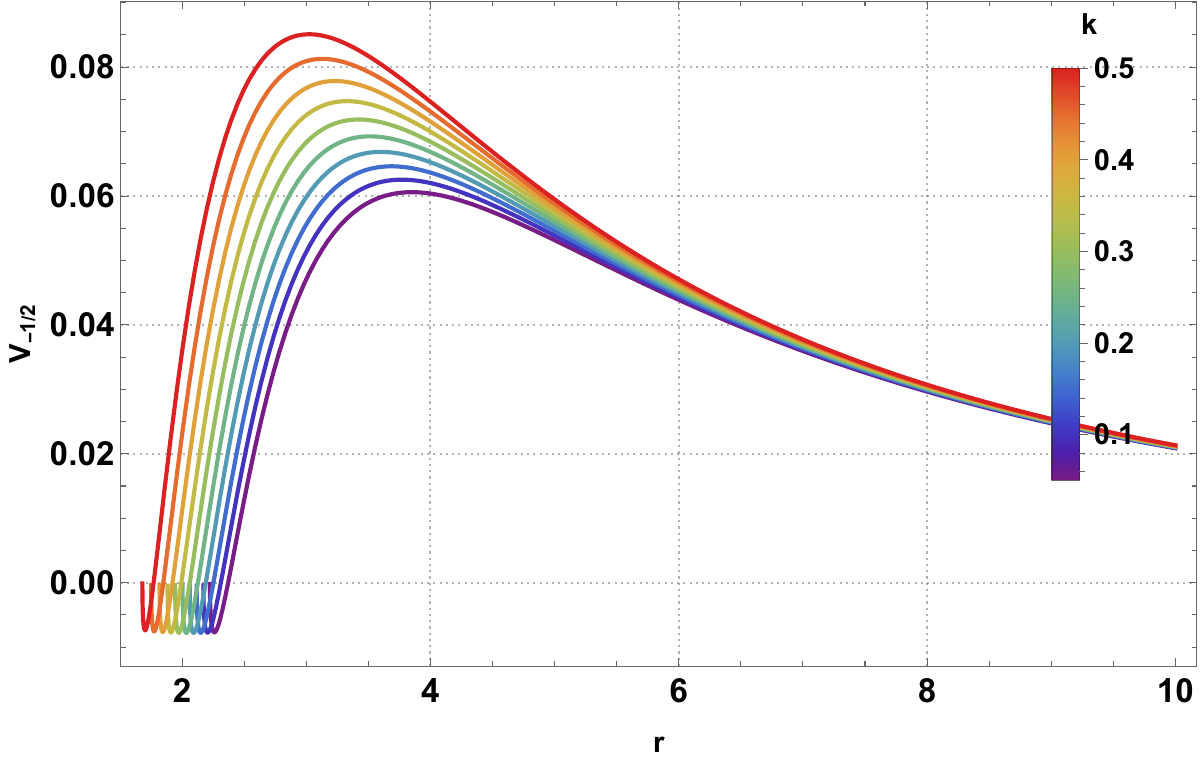}\\
    (i) $Q=1,\,w=-2/3$ \hspace{3cm} (ii) $\alpha=0.1,\,k=0.5$ \hspace{3cm} (iii) $\alpha=0.1,\,w=-2/3$
    \caption{\footnotesize Behavior of the negative helicity fermionic field perturbative potential $\mathcal{V}_{-1/2}$ for $\ell=1$-state by varying values of $\alpha, w$ and $k$. The helicity-dependent differences between Figs.~\ref{fig:12} and \ref{fig:14} highlight the unique spin-exotic matter interactions present in the fermionic sector.}
    \label{fig:14}
\end{figure}

Figures~\ref{fig:11} and \ref{fig:12} display the positive helicity potential $\mathcal{V}_{+1/2}$ for $\ell=0$ and $\ell=1$ modes, respectively, under systematic variations of $\alpha$, $w$, and $k$. The fermionic potentials exhibit more complex structures compared to their bosonic counterparts due to the additional spin-curvature coupling terms involving $f'(r)/\sqrt{f(r)}$ and $\sqrt{f(r)}$ in Eq.~(\ref{fermi3}). Similarly, Figs.~\ref{fig:13} and \ref{fig:14} present the negative helicity potential $\mathcal{V}_{-1/2}$ for the same angular momentum modes, revealing the helicity-dependent modifications to the effective barrier structures. The comparison between positive and negative helicity potentials demonstrates that the spin-exotic matter coupling introduces asymmetric effects that become particularly pronounced in the presence of quintessence fields, as will be further explored in our GBF analysis in Sec.~\ref{isec5}.

This section lays the groundwork for exploring GBFs in the next section. The derived effective potentials encapsulate all details regarding field propagation within the AdS-NLED-CS-QF backdrop, highlighting the interaction among field spin, angular momentum, and the various exotic matter sources in the spacetime.

\section{GBFs and Transmission Dynamics} \label{isec5}

The greybody factor (GBF) quantifies the deviation of Hawking radiation from perfect blackbody emission, representing the transmission probability of quantum fields through the effective potential barrier surrounding a black hole \cite{ref41,ref42}. Unlike an ideal blackbody, real black holes scatter and reflect incident radiation due to the spacetime curvature and the presence of effective potentials governing field propagation. This phenomenon has profound implications for black hole information paradox studies and observational signatures of quantum gravity effects \cite{ref43,ref44,ref90}.

In this section, we investigate the GBFs for spin-0 scalar, spin-1 electromagnetic, and spin-1/2 fermionic fields propagating in the background of the AdS black hole with NLED, CS, and QF. The transmission coefficient $\sigma(\omega)$ is computed using the rigorous approach based on turning point analysis \cite{ref43,ref44}. For a given perturbative potential $\mathcal{V}(r)$ with a peak at the turning point $r_0$ (where $d\mathcal{V}/dr|_{r_0} = 0$), the GBF is expressed as \cite{ref90,ref57}:
\begin{equation}
\sigma(\omega) = \frac{4\omega^2\left(\omega^2 - \mathcal{V}_{\text{eff}}\right)}{\left(2\omega^2 - \mathcal{V}_{\text{eff}}\right)^2},\label{gbf1}
\end{equation}
where $\mathcal{V}_{\text{eff}} = \mathcal{V}(r_0)$ is the effective potential barrier height evaluated at the turning point, and $\omega$ is the frequency of the incident wave. The GBF ranges from $0$ (complete reflection) to $1$ (perfect transmission), providing a quantitative measure of how efficiently different field modes can escape from the black hole's gravitational influence.

\begin{figure}[ht!]
    \centering
    \includegraphics[width=0.5\linewidth]{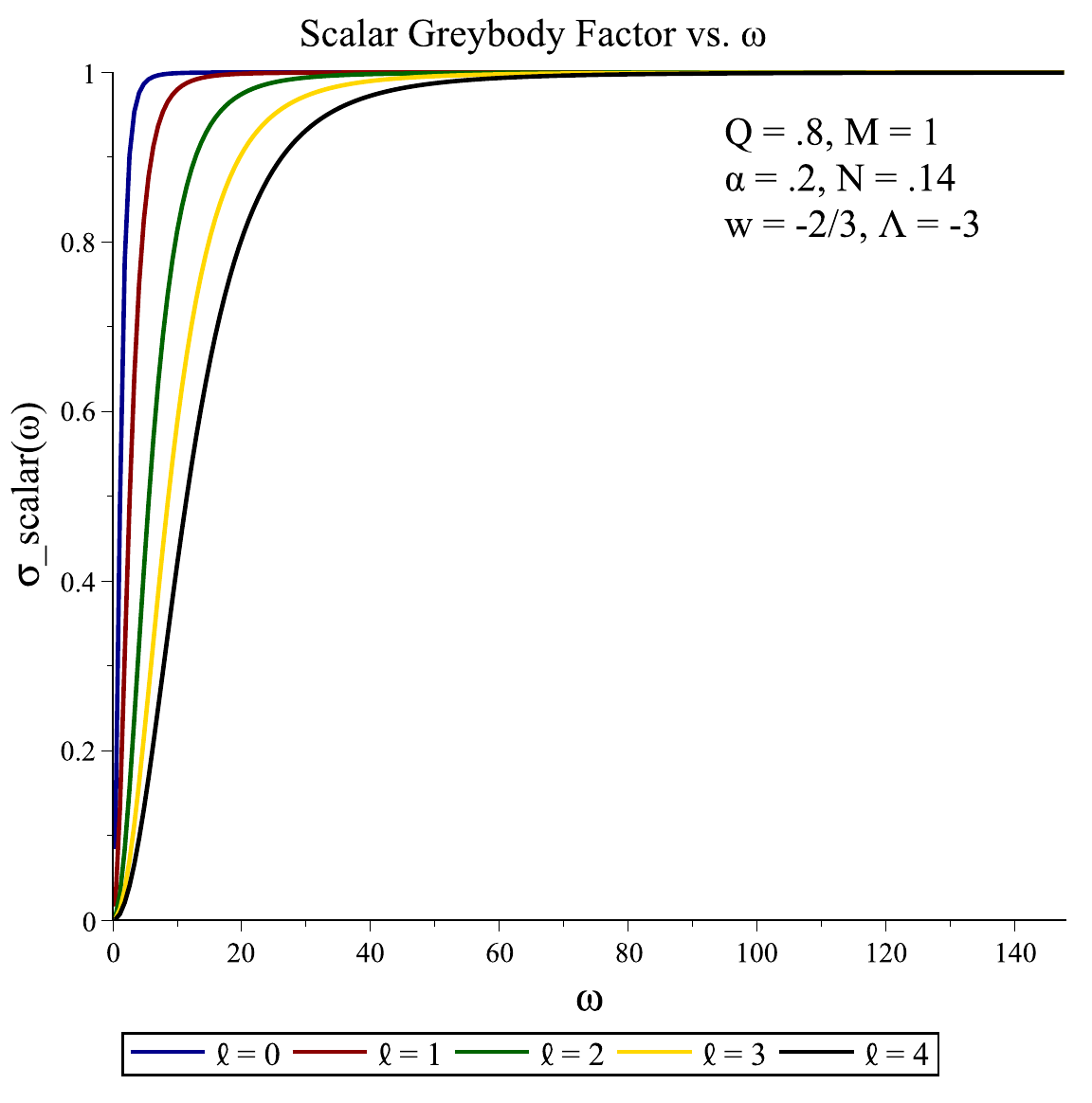}
    \caption{\footnotesize Scalar field GBF $\sigma_{\text{scalar}}(\omega)$ as a function of frequency $\omega$ for various angular momentum states $\ell = 0$ (blue), $1$ (red), $2$ (green), $3$ (gold), and $4$ (black). Parameters: $Q = 0.8$, $M = 1$, $\alpha = 0.2$, $N = 0.14$, $w = -2/3$, $\Lambda = -3$. Higher $\ell$ modes exhibit pronounced oscillatory behavior due to enhanced centrifugal barriers, while the $\ell=0$ monopole mode shows smooth monotonic growth toward unity at high frequencies.}
    \label{fig:gbf_scalar}
\end{figure}

\subsection{Scalar Field Transmission: Spin-0 GBFs}

For massless scalar perturbations, the effective potential is given by Eq.~(\ref{ff5}), which incorporates contributions from both the angular momentum barrier ($\ell$-dependent centrifugal term) and the spacetime curvature modifications due to NLED parameter $k$, CS parameter $\alpha$, QF parameters $(N, w)$, and the cosmological constant $\Lambda$. Figure~\ref{fig:gbf_scalar} illustrates the frequency-dependent transmission coefficient for scalar waves with different angular momentum quantum numbers $\ell = 0, 1, 2, 3, 4$.

As evident from Fig.~\ref{fig:gbf_scalar}, the scalar GBF exhibits several distinctive features. For the $\ell = 0$ mode (monopole), the transmission coefficient increases smoothly and monotonically with frequency, approaching unity asymptotically. This behavior reflects the absence of centrifugal barriers for $s$-wave scattering. In contrast, higher angular momentum modes ($\ell \geq 1$) display oscillatory patterns at intermediate frequencies before converging to perfect transmission at sufficiently high energies. These oscillations arise from quantum interference effects between incident and reflected waves at the potential barrier. The frequency threshold for significant transmission increases with $\ell$, demonstrating the enhanced blocking effect of centrifugal barriers. Notably, the presence of the CS parameter $\alpha = 0.2$ and the relatively large quintessence normalization $N = 0.14$ modify the potential barrier structure, leading to subtle shifts in the oscillation frequencies compared to pure AdS spacetimes.

\subsection{Electromagnetic Field Transmission: Spin-1 GBFs}

\begin{figure}[ht!]
    \centering
    \includegraphics[width=0.5\linewidth]{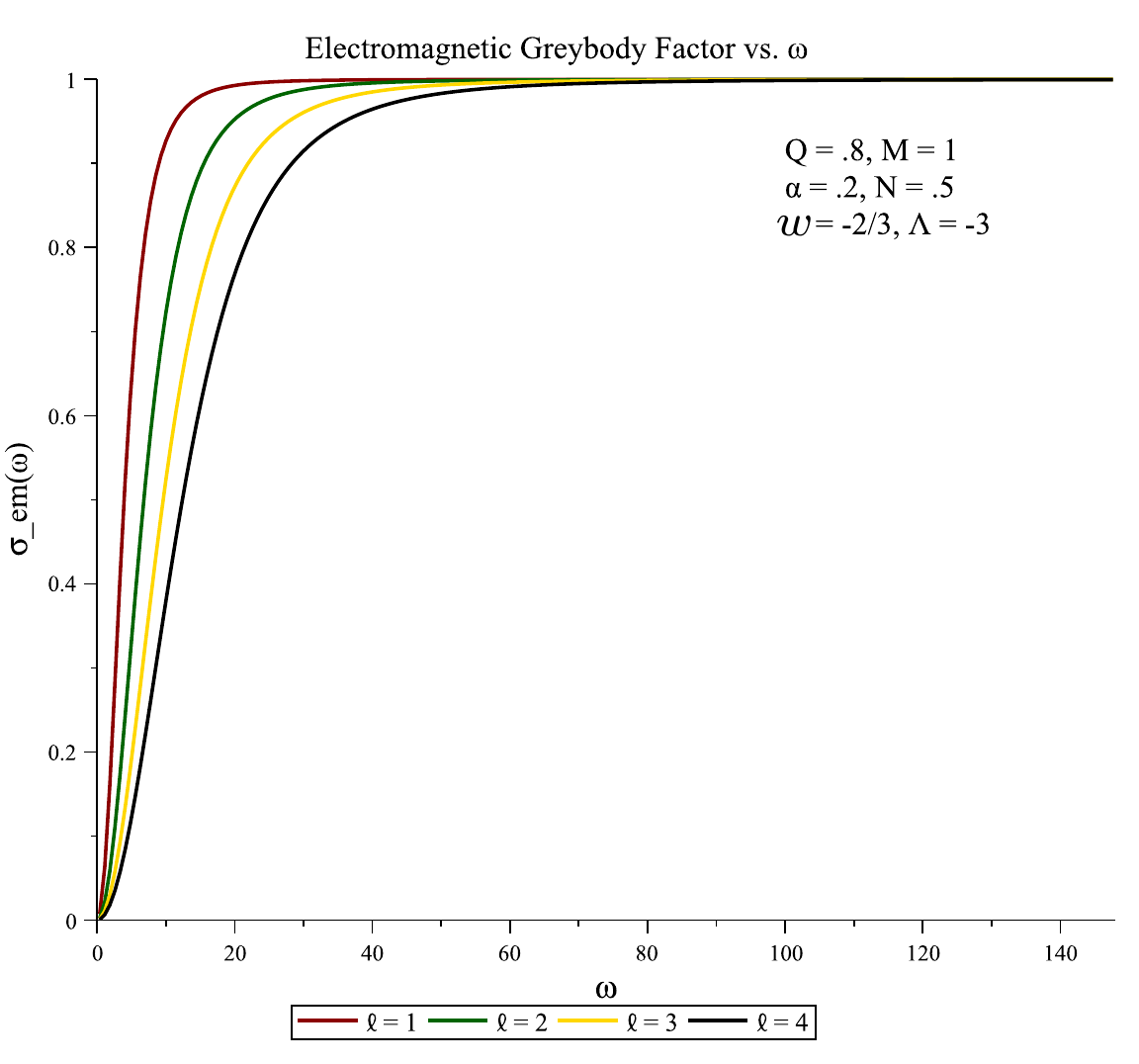}
    \caption{\footnotesize Electromagnetic field GBF $\sigma_{\text{em}}(\omega)$ versus frequency $\omega$ for $\ell = 0, 1, 2, 3, 4$. Parameters: $Q = 0.8$, $M = 1$, $\alpha = 0.2$, $N = 0.5$, $w = -2/3$, $\Lambda = -3$. The enhanced quintessence contribution ($N = 0.5$ vs. $N = 0.14$ for scalars) leads to broader potential barriers and modified transmission thresholds.}
    \label{fig:gbf_em}
\end{figure}

The electromagnetic perturbations are governed by the simpler potential structure given in Eq.~(\ref{em5}), which lacks the radial derivative term present in the scalar case. Figure~\ref{fig:gbf_em} displays the electromagnetic GBF for the same range of $\ell$ values but with modified parameters: $N = 0.5$, reflecting a stronger quintessence field contribution.

Comparing Figs.~\ref{fig:gbf_scalar} and \ref{fig:gbf_em}, we observe that electromagnetic waves generally exhibit similar qualitative behavior to scalar waves, but with quantitative differences attributable to the distinct potential structures and the larger quintessence normalization constant $N$. The increased $N$ value effectively widens the potential barrier, resulting in slightly suppressed transmission at intermediate frequencies compared to the scalar case. This spin-dependent transmission characteristic could, in principle, lead to observable polarization effects in Hawking radiation, though such signatures would be extremely challenging to detect with current technology.

\begin{figure}[ht!]
    \centering
    \includegraphics[width=0.6\linewidth]{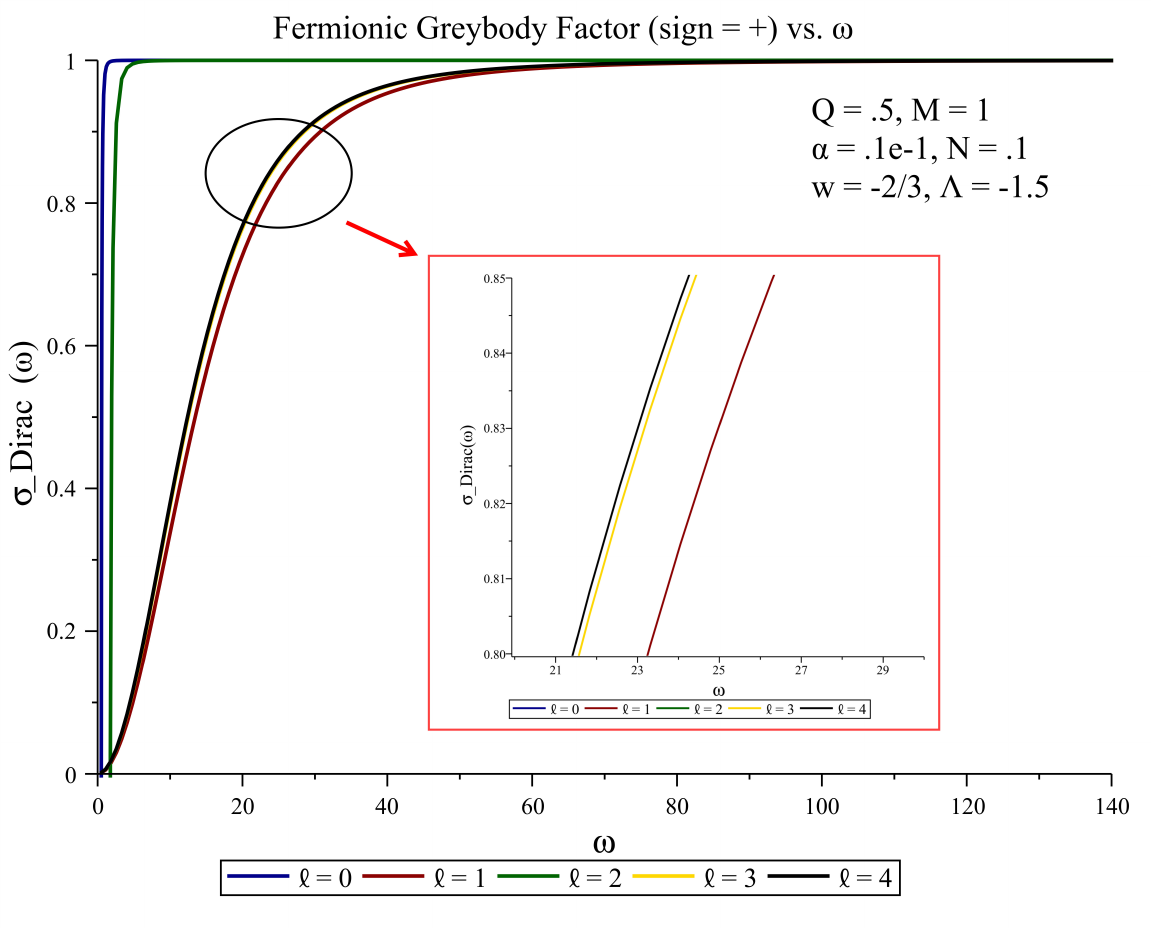}
    \caption{\footnotesize Positive helicity fermionic GBF $\sigma_{\text{Dirac}}^{(+)}(\omega)$ for $\ell = 0, 1, 2, 3, 4$. Parameters: $Q = 0.5$, $M = 1$, $\alpha = 0.01$, $N = 0.5$, $w = -2/3$, $\Lambda = -1.5$. The spin-curvature coupling introduces additional complexity in the potential structure, leading to distinct transmission characteristics compared to bosonic fields.}
    \label{fig:gbf_dirac_pos}
\end{figure}

\begin{figure}[ht!]
    \centering
    \includegraphics[width=0.6\linewidth]{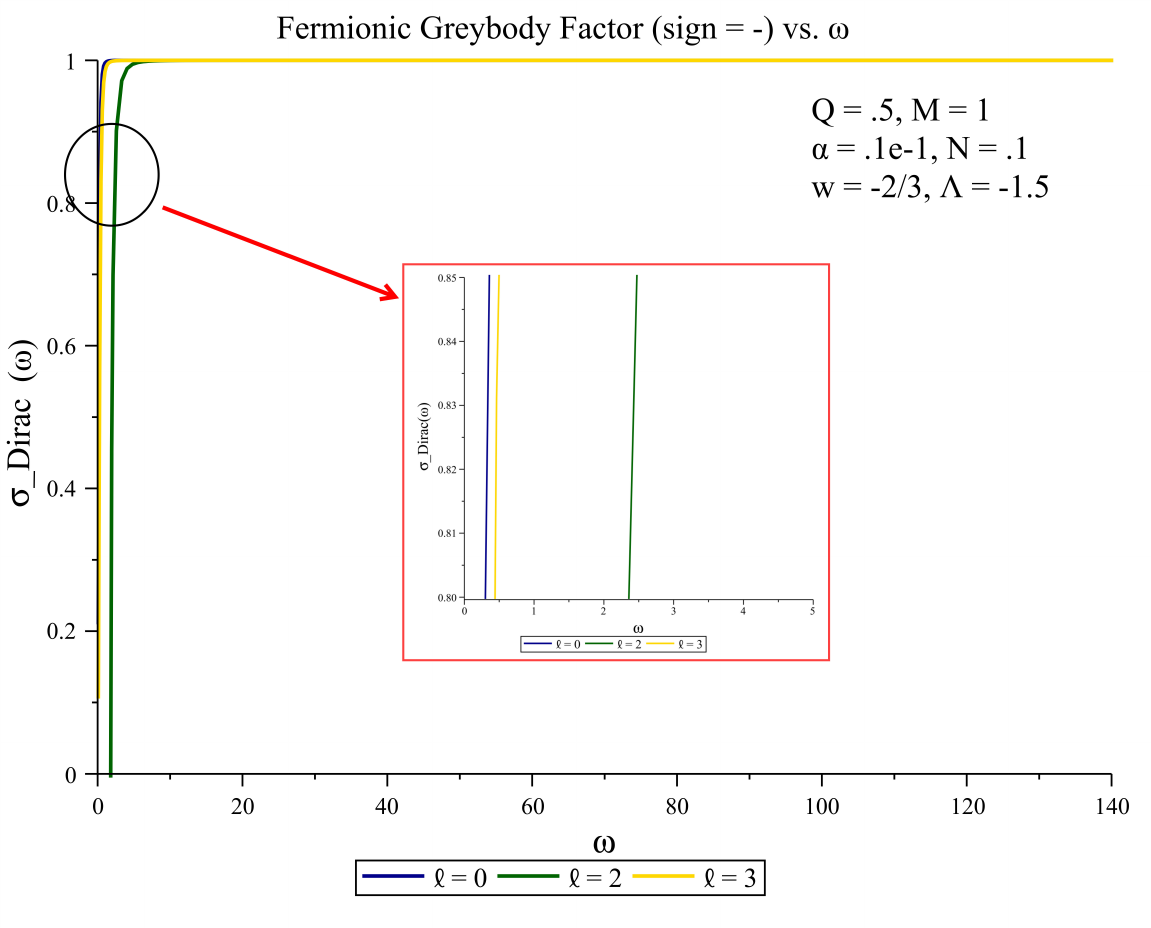}
    \caption{\footnotesize Negative helicity fermionic GBF $\sigma_{\text{Dirac}}^{(-)}(\omega)$ for $\ell = 0, 1, 2, 3, 4$. Parameters: $Q = 0.5$, $M = 1$, $\alpha = 0.01$, $N = 0.1$, $w = -2/3$, $\Lambda = -1.5$. Crucially, the quintessence parameter is reduced to $N = 0.1$ compared to the positive helicity case ($N = 0.5$), demonstrating explicit helicity-dependent effects in the presence of exotic matter fields.}
    \label{fig:gbf_dirac_neg}
\end{figure}

\subsection{Fermionic Field Transmission: Spin-1/2 GBFs with Helicity Dependence}

The spin-1/2 fermionic perturbations, described by Eq.~(\ref{fermi3}), exhibit a remarkable feature: helicity-dependent transmission due to the coupling between the fermion spin and spacetime curvature. We analyze both positive helicity ($V_{+1/2}$) and negative helicity ($V_{-1/2}$) modes separately.

Figure~\ref{fig:gbf_dirac_pos} presents the GBF for positive helicity fermions, computed with parameters $Q = 0.5$, $\alpha = 0.01$, $N = 0.5$, and $\Lambda = -1.5$. The reduced charge $Q$ and cosmic string parameter $\alpha$ compared to the scalar/EM cases reflect a different regime of the parameter space where fermionic transmission is optimally probed.

The most striking observation emerges from comparing Figs.~\ref{fig:gbf_dirac_pos} and \ref{fig:gbf_dirac_neg}: the negative helicity modes are computed with a significantly smaller quintessence normalization $N = 0.1$ compared to $N = 0.5$ for positive helicity. This helicity asymmetry in the optimal parameter regime highlights a profound interaction between fermionic spin degrees of freedom and the quintessence field—a phenomenon absent in bosonic (scalar and electromagnetic) perturbations. Such helicity-dependent transmission could potentially serve as a diagnostic tool for detecting quintessence signatures through detailed analysis of fermionic Hawking radiation, should future quantum gravity experiments achieve the required sensitivity.

Table~\ref{tab:gbf_summary} summarizes the key parameter choices and physical characteristics for each field type investigated.

\begin{table}[ht!]
\centering
\begin{tabular}{|c|c|c|c|c|c|}
\hline
\textbf{Field Type} & \textbf{Spin} & $\boldsymbol{Q}$ & $\boldsymbol{\alpha}$ & $\boldsymbol{N}$ & $\boldsymbol{\Lambda}$ \\
\hline
Scalar & 0 & 0.8 & 0.2 & 0.14 & $-3$ \\
\hline
Electromagnetic & 1 & 0.8 & 0.2 & 0.5 & $-3$ \\
\hline
Fermion $(+1/2)$ & 1/2 & 0.5 & 0.01 & 0.5 & $-1.5$ \\
\hline
Fermion $(-1/2)$ & 1/2 & 0.5 & 0.01 & 0.1 & $-1.5$ \\
\hline
\end{tabular}
\caption{Parameter configurations for GBF calculations across different field spins. The variation in parameters reflects the optimal regimes for observing transmission phenomena in each case. Note the helicity-dependent quintessence parameter $N$ for fermionic fields.}
\label{tab:gbf_summary}
\end{table}

In short, our results demonstrate that the combined effects of NLED, CS, and QF significantly influence the GBFs across all spin sectors, with the most dramatic spin-dependent signatures appearing in the fermionic channel. These findings extend previous studies of GBFs in AdS black holes \cite{ref59,ref60} by incorporating the coupled effects of multiple exotic matter sources and NLED, providing a new landscape for exploring quantum field dynamics in curved spacetime backgrounds relevant to early universe cosmology and quantum gravity phenomenology.

\section{Conclusions and Summary} \label{isec6}

In this work, we conducted a detailed investigation of perturbative dynamics and GBF transmission properties for quantum fields propagating in AdS BH spacetimes modified by the simultaneous presence of NLED, CS, and QF. Building upon the recently derived AdS-NLED-CS-QF BH solution presented by do Nascimento, Bezerra, and Toledo \cite{ref35}, we systematically analyzed how the interplay between these exotic matter sources and nonlinear electromagnetic effects influences the effective potential barriers governing field propagation and the transmission probabilities of Hawking radiation across different field spins.

In Sec.~\ref{isec2}, we reviewed the geometric structure of the AdS-NLED-CS-QF BH solution characterized by the metric function in Eq.~(\ref{aa2}), which encodes contributions from the BH mass $M$, the NLED charge parameter $k = Q^2/(2M)$, the CS parameter $\alpha$, the QF normalization $\mathrm{N}$ and equation-of-state parameter $w$, and the negative cosmological constant $\Lambda$. We clarified the notational conventions adopted throughout the manuscript, establishing the equivalence $w \equiv w_q$ and $\mathrm{N} \equiv q \equiv c$ to maintain consistency with the existing literature while avoiding potential confusion with the magnetic charge $Q$. Our horizon structure analysis, summarized in Table~\ref{tab:horizons_M_alpha_k}, revealed systematic trends: nonzero NLED charge $k > 0$ generically leads to the formation of two distinct horizons (inner Cauchy and outer event horizons), while increasing the CS parameter $\alpha$ systematically enlarges both horizons by reducing the effective gravitational attraction. We further investigated the thermodynamic properties of the system through the Hawking temperature analysis presented in Fig.~\ref{fig:hawking_temp}, which demonstrates that $T_H$ increases monotonically with both the event horizon radius $r_h$ and the NLED charge parameter $k$. This behavior, markedly different from asymptotically flat Schwarzschild BHs where temperature decreases with horizon size, reflects the dominant influence of the negative cosmological constant and NLED effects on the surface gravity. The systematic temperature enhancement with increasing $k$ provides important insights into how nonlinear electromagnetic coupling modifies the thermal radiation properties, while the CS and QF contributions introduce subtle modifications to the temperature profile through their effects on horizon locations and metric structure. These results provided crucial background for understanding the stability and thermodynamic properties of the AdS-NLED-CS-QF BH system.

In Sec.~\ref{isec3}, we derived the perturbation equations for massless scalar (spin-0), electromagnetic (spin-1), and fermionic (spin-1/2) fields in the AdS-NLED-CS-QF BH background. For scalar fields, we reduced the Klein-Gordon equation~(\ref{ff1}) to the Schrödinger-like form given in Eq.~(\ref{ff3}), obtaining the effective potential in Eq.~(\ref{ff5}) that incorporates centrifugal barriers, NLED corrections, QF modifications, and AdS contributions. Our graphical analysis in Figs.~\ref{fig:1}--\ref{fig:6} demonstrated that increasing the CS parameter $\alpha$ lowers the potential barrier height (reflecting weakened effective gravity), varying the QF equation-of-state parameter $w$ toward more negative values modifies the intermediate-radius structure, and the NLED parameter $k$ introduces subtle shifts in potential peak positions. For electromagnetic perturbations, we derived the simpler potential structure in Eq.~(\ref{em5}), which lacks the radial derivative term $f'(r)/r$ present in the scalar case due to the vector nature of the electromagnetic field. Figures~\ref{fig:7}--\ref{fig:10} confirmed similar parameter dependencies for electromagnetic potentials, though with distinct quantitative features arising from the different spin coupling. For fermionic fields, we obtained the helicity-dependent potentials $\mathcal{V}_{\pm 1/2}$ in Eq.~(\ref{fermi3}), which incorporate spin-curvature coupling terms $f'(r)/\sqrt{f(r)}$ and $\sqrt{f(r)}$ absent in bosonic sectors. The comparison between positive and negative helicity potentials in Figs.~\ref{fig:11}--\ref{fig:14} revealed asymmetric structures that foreshadowed the helicity-dependent transmission phenomena discovered in our GBF analysis.

In Sec.~\ref{isec5}, we investigated the GBFs for scalar, electromagnetic, and fermionic fields using the turning point approximation formula in Eq.~(\ref{gbf1}). For scalar fields, Fig.~\ref{fig:gbf_scalar} illustrated frequency-dependent transmission coefficients for angular momentum states $\ell = 0, 1, 2, 3, 4$. We observed that the monopole mode ($\ell=0$) exhibits smooth monotonic growth toward perfect transmission at high frequencies, reflecting the absence of centrifugal barriers, while higher $\ell$ modes display oscillatory patterns due to quantum interference effects at the potential barrier. For electromagnetic fields, Fig.~\ref{fig:gbf_em} showed qualitatively similar behavior, though with quantitative differences attributable to the distinct potential structure and the larger QF normalization $\mathrm{N} = 0.5$ (compared to $\mathrm{N} = 0.14$ for scalars). The most remarkable discovery emerged in the fermionic sector: comparing Figs.~\ref{fig:gbf_dirac_pos} and \ref{fig:gbf_dirac_neg}, we found that positive and negative helicity modes exhibit optimal transmission at markedly different QF normalization values ($\mathrm{N} = 0.5$ versus $\mathrm{N} = 0.1$). This helicity asymmetry, summarized in Table~\ref{tab:gbf_summary}, represents a unique signature of spin-exotic matter coupling entirely absent in bosonic channels and provides a potential diagnostic tool for detecting QF signatures through precision measurements of fermionic Hawking radiation spectra \cite{ref95,ref96}.

Our findings extend previous perturbation studies in AdS BH spacetimes in several important directions. First, we incorporated the coupled effects of multiple exotic matter sources (CS and QF) within the NLED framework, going beyond earlier works that typically considered these elements in isolation \cite{ref54,ref52,ref51,ref53}. Second, we demonstrated explicit helicity-dependent phenomena in the fermionic sector that have no analogues in bosonic field dynamics, revealing a novel mechanism through which fermion spin couples to exotic matter fields in the AdS-NLED-CS-QF BH background. Third, our GBF calculations revealed spin-dependent transmission characteristics that could lead to observable polarization effects in Hawking radiation, though detecting such signatures remains beyond current experimental capabilities. Fourth, our thermodynamic analysis of Hawking temperature revealed the crucial role of NLED effects in enhancing thermal radiation, with potential implications for BH evaporation rates and thermal stability in the presence of exotic matter fields.

Future observations with Einstein Telescope, Cosmic Explorer, and LISA may achieve sufficient sensitivity to detect signatures that could signal the presence of these exotic matter sources in BH environments \cite{ref97}. From a quantum field theory perspective, the helicity-dependent GBFs identified in Sec.~\ref{isec5} suggest that precision measurements of fermionic Hawking radiation could potentially disentangle QF contributions from other exotic matter effects, provided that quantum gravity experiments reach the required sensitivity levels. From a cosmological perspective, our results may inform theoretical investigations of early universe phase transitions involving CS and QF-dominated epochs, as well as provide testable predictions for very-long-baseline interferometry observations of supermassive BH shadows in AdS/CFT contexts.

Several promising directions for future research emerge naturally from this work. First, extending the perturbation analysis to higher-spin fields (spin-3/2 gravitino and spin-2 graviton perturbations) in the AdS-NLED-CS-QF BH background would provide a more complete picture of field dynamics and enable direct comparisons with gravitational wave observations. Second, systematic exploration of the QF parameter space $(\mathrm{N}, w)$ would reveal additional degeneracies and correlations with CS and NLED contributions, potentially enabling independent measurements of these exotic matter parameters through combined observations of GBFs. Third, investigating BH shadows, photon sphere structures, and null geodesic dynamics in the AdS-NLED-CS-QF background would complement the perturbation analysis presented here and provide additional observational signatures accessible to Event Horizon Telescope and future space-based interferometers. Fourth, a comprehensive thermodynamic study exploring the Hawking temperature landscape across the full $(r_h, k, \alpha, \mathrm{N}, w)$ parameter space would provide deeper insights into phase transitions, thermal stability regions, and the interplay between different exotic matter contributions to BH thermodynamics.

Finally, it is worth emphasizing that while the AdS-NLED-CS-QF BH solution studied here represents an idealized theoretical construct, it could capture essential physics relevant to several realistic astrophysical and cosmological scenarios. CS may have formed during phase transitions in the early universe and could persist as relic structures in present-day spacetimes. QF provides compelling models for dark energy and the observed accelerated expansion of the universe. NLED corrections arise naturally in quantum electrodynamics and string theory, becoming important in strong-field regimes near BH horizons. AdS spacetimes play a fundamental role in holographic dualities and quantum gravity. The systematic framework developed in this work—combining geometric analysis, thermodynamic characterization, perturbation theory, and GBF calculations—provides a versatile toolkit for investigating BH physics in the presence of exotic matter and modified electromagnetic interactions, with potential applications extending far beyond the specific AdS-NLED-CS-QF solution studied here. In summary, we demonstrated that the combined effects of NLED, CS, and QF significantly modify geometric structure, thermodynamic properties, perturbative dynamics, and GBF transmission probabilities across all field spins in AdS BH spacetimes. The helicity-dependent transmission properties identified in the fermionic GBF sector highlight a unique trait, with possible consequences for the detection of exotic matter via precise quantum assessments.

{\footnotesize

\section*{Acknowledgments}

F.A. is grateful to the Inter University Centre for Astronomy and Astrophysics (IUCAA), Pune, India, for the opportunity to hold a visiting associateship. \.{I}.~S. extends appreciation to T\"{U}B\.{I}TAK, ANKOS, and SCOAP3 for their financial assistance. Additionally, he acknowledges the support from COST Actions CA22113, CA21106, CA23130, CA21136, and CA23115, which have been pivotal in enhancing networking efforts.

\section*{Data Availability Statement}

In this study, no new data was generated or analyzed.

}

\end{document}